\documentclass[12pt]{article}

\usepackage{graphics}
\usepackage{tikz}
\usetikzlibrary{arrows}
\usepackage{amsmath}
\usepackage{amssymb}
\title{Identification of Insurance Models with Multidimensional Screening}
\author{G. Aryal, I. Perrigne \& Q. Vuong}
\date{}

\newcommand{\Real}{\hbox{\it I\hskip -2pt R}}
\newcommand{\Unit}{\hbox{\it 1\hskip -3pt I}}
\newcommand{\Complex}{\hbox{\it 1\hskip -6pt C}}
\newcommand{\Integer}{\hbox{\it I\hskip -3pt N}}
 
\begin{document}
\begin{titlepage}
\LARGE

\begin{center}
Identification of Insurance Models with Multidimensional Screening\\

\large
\vspace{0.5in}
Gaurab Aryal\\
University of Virginia\\
\vspace{0.25in}

Isabelle Perrigne\\
Rice University\\
\vspace{0.25in}

Quang Vuong\\
New York  University\\

\vspace{0.5in}
January 2016\\
\end{center}
\normalsize

\vspace{0.5in}
\normalsize
\noindent
We  thank  Pierre-Andr\'e Chiappori, Liran Einav, Matt Shum and Ken Wolpin for constructive comments.
We also benefited from participants' comments at  the Stanford Institute for Theoretical Economics,  North American Meeting of the Econometric Society and  seminars at   the Australian National University, Collegio Carlo Alberto,  Columbia University,  Georgetown University, London School of Economics,  National University of Singapore, Paris School of Economics, Sciences Politiques-Paris, Stanford University, University of Fortaleza, University of Pennsylvania,  University of Sydney and University of Wisconsin at  Madison.   
 The last two authors gratefully acknowledge financial support from the National Science Foundation through  grant SES 1148149.

\smallskip\noindent
Correspondence to   Isabelle Perrigne: iperrigne@gmail.com
  
\normalsize
\end{titlepage}

\newpage
\thispagestyle{empty}

\begin{abstract}

This paper addresses the identification of insurance models with multidimensional screening where insurees have private information about their risk and risk aversion. The model  includes a random damage and the possibility of several claims.  Screening of insurees relies on their  certainty equivalence.  The paper then investigates how data availability on the number of offered coverages and reported claims affects the identification of the model primitives under four different scenarios.
We show that the model structure is identified despite bunching due to multidimensional screening and/or a finite number of offered coverages.
The observed number of claims plays a key role in the identification of the joint distribution of risk and risk aversion. In addition, the paper derives all the restrictions imposed by the model on observables.  Our results are   constructive   with explicit equations for estimation and model testing.

\end{abstract}


\setcounter{page}{0}

\begin{center}
\maketitle
\end{center}

\section{Introduction}

Insurance has been a long studied problem in economics and is in the core of recent empirical research.  Seminal papers by Rothschild and Stiglitz (1976) and Stiglitz (1977) have provided  benchmark models of insurance under private information on insurees' risk. In  empirical studies, testing adverse selection in risk has generated a large number of papers with mixed results. See Chiaporri and Salani\'e (2000)
for the most well known test and Cohen and Siegelman (2010) for a survey of empirical findings. The recent empirical literature shows that adverse selection not only involves heterogeneity in risk but also in risk aversion, which is also called advantageous selection.   
See e.g. Finkelstein and McGarry (2006) in long-term care insurance, 
Cohen and Einav (2007) in automobile insurance, 
Fang, Keane and Silverman (2008) in health insurance, and Einav, Finkelstein and Schrimpf (2010) in annuity market. See also 
Cutler, Finkelstein and McGarry (2008) and Einav and Finkelstein (2011) for surveys.  As noted in these papers, heterogeneity in risk aversion may contradict the prediction of the benchmark adverse selection models, i.e.,
a low risk individual may buy a higher coverage because of high risk aversion and conversely. Thus, a model of insurance needs also to incorporate incomplete information in risk aversion leading to multidimensional screening. This is known to be a difficult  theoretical problem because of the violation of the Spence-Mirrlees (single-crossing)
condition. See Rochet and Stole (2003) for a survey on multidimensional screening.
 

In this paper, we propose  a  model of insurance that includes  private information in both risk and risk aversion as well as random damages and the possibility of several claims while endogenizing the contract terms. Following Landsberger and Meilijson (1999),  we consider the certainty equivalence of no insurance as a one-dimensional representation of insurees' types as this representation preserves the order of insurees after buying insurance.  For convenience, we assume a constant absolute risk aversion and a nonparametric mixture of a Poisson distribution for the number of potential claims as they lead to a tractable form for the certainty equivalence. In the spirit of the theoretical literature, we consider automobile insurance with coverages of the form premium and deductible. 
Our model contains the key ingredients of insurance and can be extended to other insurance markets such as health by adding (say) a copayment.
Thus, the model structure is defined by the joint distribution of risk and risk aversion and the distribution of damages.
Within this model,  we study the identification of the primitives.  Identification is a key step for the econometric and empirical analysis of structural models. 

Starting with Koopmans (1949) and Hurwicz (1950), the problem of identification has a long history.
As discussed by Heckman (2001),  the labor literature provides several examples of the role played by identification in empirical studies.
 Over the past fifteen years, it has received much attention with the development of structural models in empirical industrial organization. See Athey and Haile (2007) for a survey on the identification of auction models.\footnote{See also Matzkin (1994, 2007) for the nonparametric identification  of models with nonseparable errors.} 
The problem of (nonparametric) identification is important for several reasons.
First, it allows to assess the conditions required (if any) to recover uniquely the model structure from the observables while minimizing 
parametric assumptions. 
Second, it highlights which variations in the data  allows one to identify each model primitive.
Third, some important questions related to the structural analysis of models can be addressed once identification is established.
One can think of which distribution of the data can be rationalized by the model, or what restrictions the model imposes on the observables that can be used to test the model validity.

Several lessons can be drawn from 
the  recent literature on the identification of  models with incomplete information.  First, the optimal behavior of economic agents  plays an important role.  For instance, in contract models, the optimality of the offered payment  is useful in addition to the optimal agents' behavior. 
See Perrigne and Vuong (2011) for a procurement model with adverse selection and moral hazard and Luo, Perrigne and Vuong (2015) for nonlinear pricing. Thus, in most cases  we need to consider both sides of the market, i.e. the principal and the agent(s), and assume that the observations are the equilibrium outcomes.
Second, one achieves  identification with standard identifying strategies such as instrumental variables and  exclusion restrictions,  that  have been widely used in the early literature on identification.  
See Guerre, Perrigne and Vuong (2009) for the identification of risk aversion in auctions and Berry and Haile (2014) for a recent contribution to the identification of multinomial choice demand models.
Third, the one-to-one equilibrium  mapping between the unobserved agent's private information and the observed outcome  
 is a key element on which identification relies.
See e.g. Guerre, Perrigne and Vuong (2000) and Athey and Haile (2007) in the context of auctions.

Our paper differs from this literature in several dimensions. 
First, we consider a model with multidimensional screening in which  bunching/pooling cannot be avoided. In this case, identification cannot rely exclusively on the one-to-one mapping between the agent's unobserved types and his observed 
outcome/action.\footnote{Relying on Rochet and Chone (1998), Pioner (2007)  addresses the semiparametric identification of bidimensional screening  models in a nonlinear pricing context but  assumes that one of the two agent's types is observed by the analyst.  Aryal (2015)  considers nonparametric identification with multidimensional types.  See also 
Luo, Perrigne and Vuong (2012, 2013) who study identification of nonlinear pricing models with multiple types relying on  Armstrong (1996) model. The latter papers use optimality of both the principal and the agent as well as observations from multiple markets to identify the model primitives.}
Second, our model also  considers the possibility of  a finite number of options/contracts offered to each agent, while agents' types are distributed over a continuum.
In addition to the bunching arising from multidimensional screening, additional bunching arises
because of a finite number of contracts.  This represents an additional challenge in the study of identification.\footnote{
Crawford and Shum (2007) consider two contracts while agents' types  can take only two values  thereby avoiding any bunching. Gayle and Miller 
(2015) adopt a similar strategy. Leslie (2004) entertains a finite number of price options through a discrete choice model to analyze consumers' behavior
taking the price schedule as exogenous.}

To study identification of the model primitives and assess how data availability affects identification,  we proceed as follows. We consider several data scenarios  depending on the number of offered coverages and reported claims, namely whether the  number of coverages is a continuum or finite and whether the claims contain all the information, or only those above the deductible. 
This strategy  allows us to assess how data constraint or limit the identification of primitives, and which identifying assumptions  are needed.  Moreover,
studying the identification under a continuum of coverages is important as a negative identification result would imply nonidentification of the model primitives under a finite number of coverages.
A first data scenario  exploits the one-to-one mapping between the level of certainty equivalence and the deductible to identify the distribution of certainty equivalence. The number of claims then plays a crucial role in identifying  the joint  distribution of risk and risk aversion.  
A second data scenario maintains a continuum of contracts but considers a damage distribution truncated at the deductible. Because 
a continuum of contracts is offered, the subpopulation choosing full insurance, i.e., a zero deductible, identifies the damage distribution and the argument of the first case applies.

When considering a finite number of contracts in the third and the fourth scenarios, identification becomes more challenging as  we cannot exploit a one-to-one mapping between 
(say) the deductible and the insuree's private information. Though the context is different, the number of claims continues to play a key role in identifying the marginal distribution of risk.   Regarding the identification of the joint distribution of risk and risk aversion, we exploit an exclusion restriction and a full support assumption requiring  sufficient variations in some exogenous characteristics.
Under these assumptions, the model structure is identified when the damage distribution is fully observed.
On the other hand, when the damage distribution is truncated at the deductible, we obtain identification of the structure up to the knowledge of the probability that the damage is below the deductible. The latter probability is not identified. We then 
discuss some identifying assumptions for the probability of damage below the deductible.  A notable feature  of our results under a finite number of contracts is that they  do not rely on the optimality of the offered coverages. Consequently, our results apply to any form of competition in the insurance industry. 
To complete these results, we derive all the model restrictions on the observables in the fourth data scenario.  These restrictions can be used to test the validity of the model and its assumptions.  For instance, a model restriction allows a test of  optimality  of the offered coverages.
This contrasts with the previous literature as discussed above, and our results represent a novel perspective to the identification of models under incomplete information. In addition, 
all our results are constructive and provide explicit equations for estimation and testing.

The  paper is organized as follows. Section 2 presents the model.  Sections 3 and 4 study identification under a continuum of contracts  and under a finite number of contracts, respectively.
Section 5 discusses some identifying strategies for the damage probability below the deductible and  derives all the restrictions imposed  by the model on observables.
Section 6 concludes with future lines of research. An
 appendix collects the proofs.

\section{A  Model of Insurance  }

This section develops a model in which insurees have private information about their  risk and risk aversion. The presence of multiple private information leads to multidimensional screening with pooling at equilibrium. See  Rochet and Stole (2003) for a survey.  Following Landsberger and Meilijson (1999),  we use the concept of certainty equivalence  to rank and  screen insurees.  To fix ideas and  in the spirit of the early literature, we consider automobile insurance as an example throughout the paper though our framework  also applies to (say) homeowner and rental insurance. See  the end of this section for a discussion of health insurance.

\bigskip\noindent
{\sc The Benchmark Model by Stiglitz}

This section briefly reviews the Stiglitz (1977) model and motivates our  model that incorporates heterogenous preferences and a random damage/expense. It also introduces  basic notations.
Insurees are characterized by a probability of accident (risk) $\theta \in [\underline{\theta},\overline{\theta}]$ distributed as $F(\cdot)$ with a density $f(\cdot)$. An accident involves a fixed damage $D$ affecting the insuree's wealth $w$. Because agents are risk averse,   they  buy  insurance  by paying a premium $t$. The insurance company requires a deductible $dd$ for each accident. Upon buying insurance, the agent's wealth is $w-t$ in the event of no accident with probability 
$1-\theta$ and $w-t-D+(D-dd)=w-t-dd$  in the event of an accident with probability $\theta$. His expected utility is then 
$V(t,dd;\theta)= (1-\theta)U(w-t)+\theta U(w-t-dd)$, where $U(\cdot)$
is a von Neumann-Morgenstern utility function, which is continuous, strictly increasing and concave. The risk $\theta$ is private information while $U(\cdot)$ is known by the insurer.

In an incomplete information setting, the insurance company offers contracts of the form $[t(\theta),dd(\theta)]$ that are incentive compatible. The firm's profit from a $\theta$-insuree is 
$\pi(\theta)= (1-\theta) t(\theta)+\theta[t(\theta)-D+dd(\theta)]= t(\theta)-\theta (D-dd(\theta))$.  Because risk is unknown, the insurance company maximizes its expected profit  subject to
the insuree's incentive compatibility (IC) and participation (IR) constraints, namely 
\begin{eqnarray*}
&&\max_{t(\cdot),dd(\cdot)}  \int_{\underline{\theta}}^{\overline{\theta}}
\pi(\theta) f(\theta) d\theta \\
&&{\rm s.t.} \ \ - t'(\theta)- dd'(\theta)= \frac{1-\theta}{\theta} t'(\theta) \frac{U'(w-t(\theta))}{U'(w-t(\theta)-dd(\theta))} \ {\rm (IC)}\\
&&  \qquad V(t(\theta),dd(\theta);\theta) \geq  (1-\theta)U(w)+ \theta U(w-D) \ \ \ \quad\  {\rm (IR)},
\end{eqnarray*}
where the RHS of  (IR) expresses the agent's expected utility with no insurance.\footnote{Because $U(w)-U(w-D)>0$, there is no countervailing incentives as defined by Lewis and Sappington (1989).}

The main findings of this model are as follows.  First, pooling is not optimal and the firm benefits from offering a continuum of contracts. The individual with the highest risk, i.e., $\overline{\theta}$, is offered full insurance with a zero deductible.
Second, premium and deductible are inversely related. In addition,  the premium is a convex  function of the deductible implying  a larger marginal price for lower deductibles.  
Third, the optimal coverage may entail some optimal exclusion for  insurees with a low probability of accident. 

Though insurance contracts can include several features such as copayments and hard limits, it is worth noting that Arrow (1963) shows the optimality of the premium-deductible contract.  Intuitively,  the latter allows the best risk-sharing between a risk neutral insurer and a risk averse insuree as it is the best compromise between the willingness to reduce risk and the need to limit the insurance deadweight cost. Furthermore, Gollier and Schlesinger  (1993) show that any other form of insurance contract is dominated  by a contract  with a deductible and a premium implying that the deductible-premium coverage maximizes insurer's profit over all other possible forms of implementable contracts.

The above model assumes at most one accident with  a fixed damage and the same (known) risk aversion across insurees. In reality, there might be more than one accident over the policy period and every accident involves a random damage. Moreover, as shown by Finkelstein and McGarry (2006) and  Cohen and Einav (2007), the variability in risk aversion might be more important than the variability in risk across insurees.  It is also natural to consider that insuree's risk aversion is as private as his probability of accident.  Consequently, asymmetric information becomes bidimensional. Ignoring heterogeneity in risk aversion may have serious consequences on insurance policy design.
For instance,  an insuree with a low probability of accident and a high risk aversion may buy a contract with a high level of coverage (or low deductible) and conversely. This is also known as advantageous selection in the insurance literature.  
 In contrast, when heterogeneity in  risk aversion is ignored as in the above model, this insuree should buy a low level of coverage.
In addition, the distribution of damages as well as the expected number of accidents have an important impact on the choice of deductible relative to the premium offered by the insurer.  In view of this discussion, our model includes multiple accidents with random damage and heterogeneity in privately known  risk aversion. In view of data availability,  our model also considers a finite number of offered contracts/coverages.

\bigskip\noindent
{\sc  Model Assumptions}

We make the following assumptions. In our model, $\theta$ is the insuree's risk measured as the expected number of accidents over the period of coverage.

\medskip\noindent
{\bf Assumption A1:}{\em

\noindent
(i) The insuree's utility function exhibits Constant Absolute Risk Aversion (CARA),  i.e., $U(x;a)= - \exp(-ax), a>0$,

\noindent
(ii) The pairs $(\theta,a)$  are i.i.d. as $F(\cdot,\cdot)$ which is twice continuously differentiable on
 its support  $\Theta \times {\cal A}= [\underline{\theta},\overline{\theta}] \times [\underline{a},\overline{a}] \subset \Real_{++} \times \Real_{++}$, 

\noindent
(iii)  Each insuree may be involved in $J$ accidents, which  conditional on $\theta$, follows a 
Poisson distribution, i.e. $p_j(\theta)={\rm Pr}[J=j|\theta]= e^{-\theta} \theta^j/j!$,

\noindent
(iv) $J$ is independent across insurees and each accident involves a damage $D_j$, 
$j=1,\ldots,J$. The damages  are i.i.d as $H(\cdot)$ on support 
$[0,\overline{d}] \subset \Real_+$,

\noindent
(v)  $D_j, j=1,\ldots,J$ is independent of $(\theta,a)$.

}

\medskip
By A1-(i), the utility function is strictly increasing and concave. The CARA specification has two main advantages: (i) It leads to a tractable expression for the certainty equivalence and (ii) the attitude toward risk in changes in wealth is independent of initial wealth. These properties have made the CARA utility a popular choice in the theoretical and empirical literature.
By A1-(ii), each insuree is characterized by a pair $(\theta,a)$ which is private information.
Assumption A1-(iii) specifies the distribution of accidents as Poisson with mean $\theta$. This distribution is widely used in actuarial science to model 
the number of accidents. The combination of the CARA utility and the Poisson distribution is especially convenient as it leads to explicit expressions for the certainty equivalence defined later. Since $\theta$ is random by 
A1-(ii), and its marginal distribution is left unspecified, the distribution of the number of accidents in the population is a nonparametric mixture of Poisson distribution thereby  adding  flexibility.\footnote{Cohen and Einav (2007) consider a  log normal  mixture of Poisson  distribution.} Relaxing the CARA and/or Poisson specifications is possible at the cost of obtaining implicit expressions for the certainty equivalence.
Our identification results of Section 3 and 4 would still hold provided the distribution of the number of accidents belongs to the  class of distributions whose nonparametric mixture is identified. See Rao (1992).
By A1-(iv,v), damages are random, mutually independent and independent of types $(\theta,a)$.  We view the damage as being affected by exogenous factors such as bad luck, weather or road conditions. Its independence with
$(\theta,a)$ excludes moral hazard  as (say) risk averse agents' action might reduce the damage per accident.  This issue  is left for future research.  Section 5.2 discusses how A1-(iv,v) can be tested in view of the restriction it implies on observables.

Lastly, following Stiglitz (1977) and empirical papers such as Cohen and Einav (2007) among others, we assume the insurer acts as a monopolist.  The concentration ratios and the profits made in the insurance industry indicate that it is not a competitive market. See  Chiappori, Julien, Salanie and Salanie (2006) for automobile insurance and Dafny (2010) and Starc (2014) for health insurance.  For instance, switching costs for  automobile and home  insurance and/or the limited number of employer offered coverages in health insurance may prevent  insurees to benefit from competition. See Israel (2005a,b) and Honka (2014) for evidence in the automobile industry.
Considering an oligopoly  would add great complexity to the model because of  the  increasing dimension of adverse selection  due to product differentiation. In view of this, we consider the monopoly as a reasonable trade-off.

The model primitives $[F(\cdot,\cdot), H(\cdot)]$ are common knowledge.
The timing is as follows.  Each insuree draws independently  a pair of types $(\theta,a)$ from $F(\cdot,\cdot)$. The insurance company proposes a menu of  insurance contracts of the form $[t,dd]$, where $dd$ is the deductible per accident. The insuree chooses the contract that maximizes his utility and pays the corresponding premium.
In case of an accident with damage below the deductible, the insuree pays for it.
Otherwise, the insurer pays the damage above the deductible and the insuree pays the deductible.  

\bigskip\noindent
{\sc Insurer's Optimization Problem}

The insurer offers a continuum of contracts $[t(\theta,a),dd(\theta,a)]$
for $(\theta,a) \in \Theta \times {\cal A}$.  Integrating over $(\theta,a)$, the insurer's expected profit is given by
\begin{eqnarray}
{\rm E}[\pi(\theta,a)] 
&=& \int_{\Theta\times{\cal A}} \left[t(\theta,a)- \theta \int_{0}^{\overline{d}}\max\{0,D-dd(\theta,a)\}  dH(D)\right] dF(\theta,a) \nonumber\\
&=& \int_{\Theta \times {\cal A}} \left[ t(\theta,a) - \theta \int_{dd(\theta,a)}^{\overline{d}}(1-H(D)) dD \right] dF(\theta,a),
\end{eqnarray}
where $\max\{0,D-dd(\theta,a)\}$ reflects that the insurer only covers the damage above the deductible. 
The first equality uses that  damages are i.i.d  conditional on $(\theta,a)$ by A1-(iv,v) while  the second equality follows from integration by parts. The inside integral is the expected payment per accident while $\theta$ is  the expected number of  accidents.

For a $(\theta,a)$-individual with wealth $w$, his  expected utility without insurance is
\begin{eqnarray}
V(0, 0;\theta,a)&=& p_0(\theta) U(w;a)+p_1(\theta){\rm E}[U(w-D_1;a)]+ p_2(\theta) {\rm E}[U(w-D_1-D_2;a)]+\ldots \nonumber\\
                 &=& -p_0(\theta) e^{-aw}-p_1(\theta)e^{-aw}{\rm E}[e^{aD_1}]-p_2(\theta) e^{-aw} {\rm E}[e^{aD_1}]{\rm E}[e^{aD_2}]-\ldots\nonumber\\
                 &=&-e^{-aw}\left[p_0(\theta)+p_1(\theta)\phi_a +p_2(\theta) \phi_a^2+\ldots \right]\nonumber\\
                  &=& -e^{-aw} e^{-\theta}\left( 1 + \frac{\theta \phi_a}{1!} + \frac{\theta^2\phi_a^2}{2!}+\ldots\right)\nonumber\\
                  &=& -e^{-aw+\theta(\phi_a-1)},
\end{eqnarray}
where $\phi_a={\rm E}[e^{aD}]>1$, and the expectation is with respect to $D$. The first equality  considers all the possibilities regarding the number of accidents and their costs to an individual without insurance.   The second equality uses the CARA utility function  and the independence of damages across accidents by A1-(i,iv,v).
The third equality uses damages being identically distributed by A1-(iv).
Lastly, the fourth equality relies on  the Poisson distribution of accidents by A1-(iii).  
Using the same derivation where $w$ and $D_j$ are replaced by $w-t$ and 
$\min\{dd,D_j\}$, respectively, the expected utility  of a 
$(\theta, a)$-individual  buying insurance $(t,dd)$ is
\begin{eqnarray}
V(t,dd;\theta,a)= -e^{-a(w-t)+\theta\left(\phi_a^*-1\right)}, 
\end{eqnarray}
where $\phi_a^*={\rm E}[e^{a\min\{dd,D\}}]= \int e^{a\min\{dd,D\}}dH(D)=\int_{0}^{dd} e^{aD}dH(D)+e^{add}(1-H(dd))>1$.
We remark that $\phi_a^* < \phi_a$ as $\min\{dd,D\} \leq D$.

Given a menu of contracts, the $(\theta,a)$-individual chooses the contract that maximizes his expected utility as defined above. Following the revelation principle, we can focus on  a direct mechanism that maps types
to contract terms, i.e. $[t(\theta,a),dd(\theta,a)]$. The insurer, however, should choose implementable contracts that satisfy the insuree's optimization or (IC) constraint as well as the insuree's participation or (IR) constraint. This gives the following optimization problem
\begin{eqnarray}
&&\max_{t(\cdot,\cdot),dd(\cdot,\cdot)} {\rm E}[\pi(\theta,a)]\\
&& \ \ {\rm s.t.}\   V[t(\theta,a),dd(\theta,a);\theta,a] \geq V[t(\tilde{\theta},\tilde{a}),dd(\tilde{\theta},\tilde{a});\theta,a]\ \ \forall (\tilde{\theta},\tilde{a}) \in \Theta \times {\cal A}\qquad  (IC) \nonumber \\
&& \qquad \ V[t(\theta,a),dd(\theta,a);\theta,a]  \geq V(0,0;\theta,a)\ \  \forall 
(\theta,a) \in \Theta \times {\cal A} \qquad \qquad \qquad\ \  (IR), \nonumber
\end{eqnarray}
where the expected profit is given in (1). The (IC) constraint ensures that the $(\theta,a)$-individual chooses the contract $(t(\theta,a),dd(\theta,a))$.
The (IR) constraint guarantees that buying this contract is  better for this individual  than having no insurance.

As is well known, multidimensional types leads to a complex screening problem. See Rochet and Stole (2003). As noted previously, an insuree with high risk but low risk aversion might have the same willingness to pay for a given coverage $(t,dd)$ as an insuree with low risk but high risk aversion. This substitutability between risk and risk aversion implies that a separating equilibrium, where each individual $(\theta,a)$ gets a unique coverage, is infeasible. Thus, pooling occurs across insurees.  Intuitively,  insurees have two sources of private information while the insurer has in fact a single instrument, the deductible, to screen insurees.
Indeed, the premium and deductible are inversely related as  a contract $(t,dd)$ will be always preferred to any other contract $(t,dd')$ with 
$dd' > dd$.  
Thus, the insurer's objective is to find  the best way to pool insurees such that  offered coverages are feasible, i.e., satisfy the (IC) and (IR) constraints, while maximizing its expected profit.\footnote{A simple argument shows that screening on risk or risk aversion only is not optimal for the insurer.  Consider three individuals $(\theta_1,a_1)$, $(\theta_2,a_1)$ and $(\theta_2,a_2)$ with $\theta_1<\theta_2$ and $a_1<a_2$, then having (say) the first and second buying the same coverage is not optimal for the insurer's  profit.}  

\bigskip\noindent
{\sc Certainty Equivalence}

Following Landsberger and Meilijson (1999), we use the certainty equivalence of no insurance as a one-dimensional  aggregation of  the two dimensions of private information.
A similar aggregation approach was proposed by Laffont, Maskin and Rochet (1987).\footnote{See also Ivaldi and Martimort (1994) for an application to competitive nonlinear pricing. For alternative approaches, see (say) Wilson (1993)  who adopt a partitioning of the types set into one-dimension subsets, Rochet and Chone  (1998) who propose a general approach for multidimensional screening when the number of types is equal to the number of instruments, and Basov (2001) for  the general case.} Screening based on certainty equivalence has two main advantages. First, it does not rely as much on parametric specifications of the model primitives. Second, certainty equivalence has a natural economic interpretation. See also Armstrong (1996) who uses the  production cost for a multiproduct firm to screen consumers with multidimensional types. We make the following assumption.

\medskip\noindent
{\bf Assumption A2:} {\em  For any given coverage $(t,dd)$, the difference $V(t,dd;\theta,a)-V(0,0;\theta,a)$ is increasing in $a$. 
}

\medskip
We remark that the above difference  is automatically increasing in $\theta$.
Thus, individuals with  higher risk or risk aversion value  insurance more than those with lower risk or risk aversion.  Assumption A2  ensures that there will be no countervailing incentives because  the (IR) constraint in (4), namely $V(t,dd;\theta,a)-V(0,0;\theta,a) \geq 0$ has a LHS increasing in both $(\theta,a)$.  We note that A2 restricts the coverage $(t,dd)$ for a $(\theta,a)$-individual  relative to  the damage distribution. This assumption can be verified ex-post upon identification of the model primitives.

The certainty equivalence  $CE(0,0;\theta,a)$ of no insurance coverage  is defined by the amount of certain wealth for the insuree that will give him the same level of utility when he has no coverage, i.e.,  $-\exp(-aCE(0,0;\theta,a))=V(0,0;\theta,a)$. Thus, by (2) 
\begin{eqnarray}
 CE(0,0;\theta,a)=w-\frac{\theta(\phi_a-1)}{a}.
\end{eqnarray}
The certainty equivalence $CE(t,dd;\theta,a)$ of having the coverage $(t,dd)$ is defined similarly as the amount of certain wealth for the insuree that will give him the same level of utility when buying coverage, i.e. 
$-\exp(-aCE(t,dd;\theta,a))=V(t,dd;\theta,a)$. Thus, by (3)
\begin{eqnarray}
CE(t,dd;\theta,a)=w-t-\frac{\theta(\phi^*_a-1)}{a}.
\end{eqnarray}
The next lemma establishes the monotonicity in $(\theta,a)$ of these certainty equivalences. All proofs are in the Appendix.

\medskip\noindent
{\bf Lemma 1:} {\em The certainty equivalences  (5) and (6) are both decreasing in risk and risk aversion.}

\begin{figure}[t!]
\centering
\includegraphics[scale=0.5]{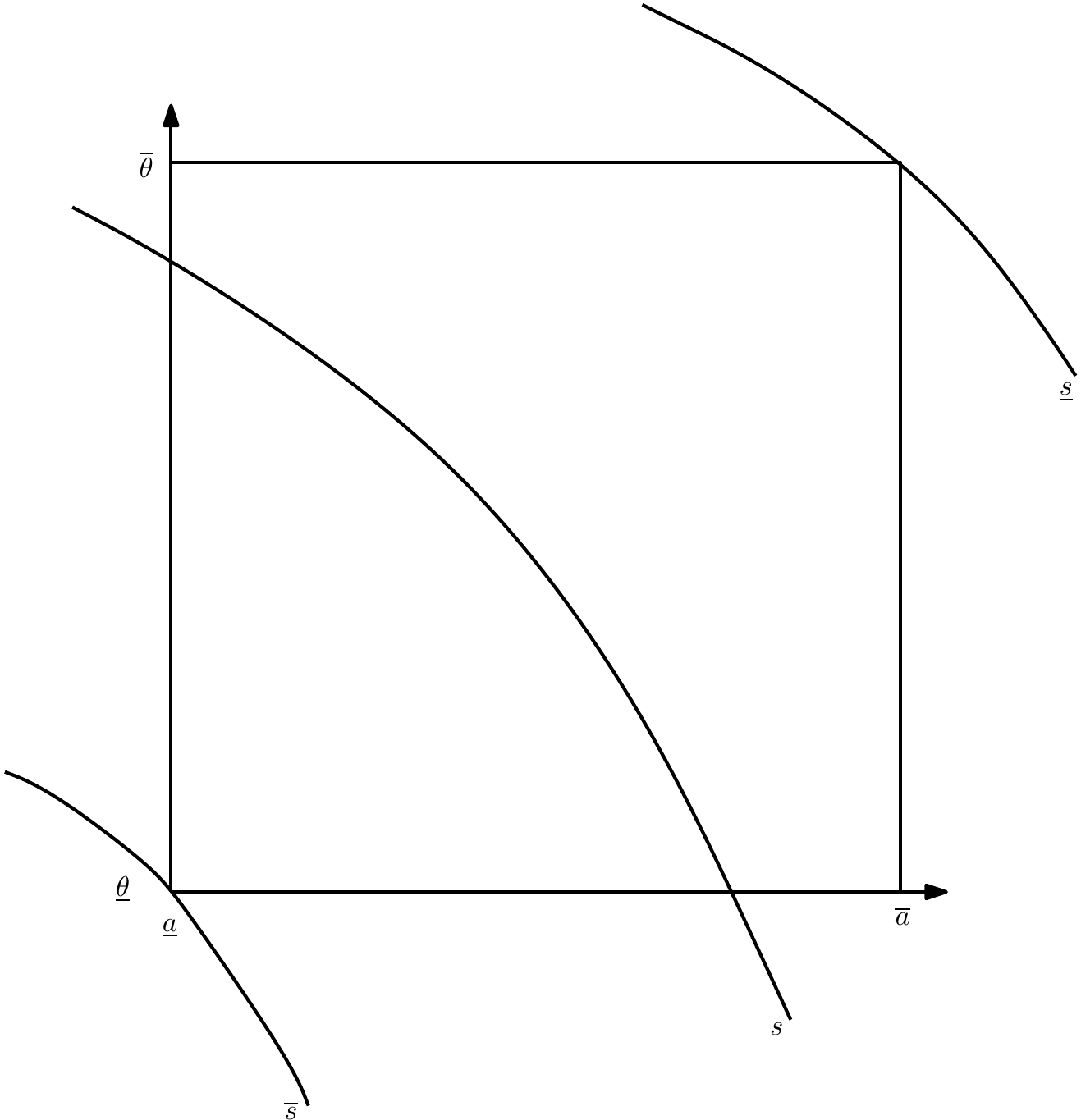}
\caption{Certainty Equivalence\label{CE}}
\end{figure}

\medskip
The certainty equivalence of no insurance in  (5)  defines a locus of pairs $(\theta,a)$ as a downward sloping curve   $\theta(a)$ for
any  given value $s$ of certainty equivalence.
Because $s\equiv CE(0,0;\theta,a)$ is a function of $(\theta,a)$, namely $s(\theta,a)$, it is random and  distributed as $K(\cdot)$ with some density $k(\cdot)$ on $[\underline{s},\overline{s}]$, 
where $\underline{s}=s(\overline{\theta},\overline{a})$ and 
$\overline{s}=s(\underline{\theta},\underline{a})$, respectively.
 Figure \ref{CE} displays some $s$-isocurves.

\bigskip\noindent
{\sc Solving the Multidimensional Screening Problem}

The optimization problem (4) is known to be difficult to solve because of  multidimensional private information and the loss of the single-crossing property. The literature on multidimensional screening shows that pooling at equilibrium cannot be avoided. 
To make the parallel with the literature on multidimensional screening with a  focus on nonlinear pricing, we remark that the premium $t$ plays the role of the payment  and $-dd$ plays the role of the quantity as seen in (3) and
(6). Thus, $dd$ is   the only  instrument  for two dimensional types.  
The certainty equivalence without insurance aggregates these two dimensions into a single one.
We then rewrite the optimization problem (4) in terms of $s\equiv CE(0,0;\theta,a)$ and the (IC) and (IR) constraints  using the certainty equivalences $CE(t,dd;\theta,a)$
and $CE(0,0;\theta,a)$.   Cohen and Einav (2007) also use the certainty equivalence to explain the choice of coverage by insurees.

Here, we consider a continuum of contracts $[t(s),dd(s)]$   for $s \in [\underline{s},\overline{s}]$.
 Thus all  insurees with the same value of certainty equivalence $s$ are pooled. 
 Intuitively, the insuree with the highest outside option or the highest certainty equivalence will be  treated as the individual with the lowest willingness-to-pay or the lowest risk individual in  Stiglitz (1977). The insurer needs to propose an attractive coverage with a high deductible to induce truth-telling and participation because he values insurance the least.  On the other hand, the individual with the lowest outside option or the lowest certainty equivalence is offered full coverage  or $dd=0$ as shown later. 
Landsberger and Meilijson (1999) show that optimal  insurance contracts preserve the order of certainty equivalence, i.e.,
for any pair of types $(\theta,a)$ and 
$(\theta',a')$ such that $s(\theta,a) < s(\theta',a')$ or $s<s'$, the optimal contract $[t(s),dd(s)]$ satisfies $s(\theta,a) \leq CE(t(s),dd(s);\theta,a) < s(\theta',a') \leq CE(t(s'),dd(s');\theta',a')$.
The  first and third inequalities come from the (IR) constraints, i.e., an individual will buy insurance if his utility is larger than not buying insurance.
This property  ensures that screening on $s$ is implementable.

We rewrite the expected profit (1) in terms of $s$.  Noting  $t(\theta,a)=t(s)$ and $dd(\theta,a)=dd(s)$ and making the change of variables $(\theta,a)$ to $(\theta,s)$ in (1)  give   
\begin{eqnarray}
{\rm E}[\pi]=\int_{\underline{s}}^{\overline{s}}\left[t(s)-{\rm E}(\theta|s)\int_{dd(s)}^{\overline{d}}(1-H(D))dD\right]k(s)ds,
\end{eqnarray}
where $k(\cdot)$ is the density of certainty equivalence $s$.  The (IC) and (IR)
constraints in (4) become
\begin{eqnarray*}
&&CE(t(s),dd(s);\theta,a) \geq CE(t(\tilde{s}),dd(\tilde{s});\theta,a) \ \forall 
\tilde{s} \in [\underline{s},\overline{s}] \qquad \qquad\qquad \qquad\qquad \quad \ \ \ (IC) \\
&& CE(t(s),dd(s);\theta,a) \geq s, \qquad\qquad\ 
\qquad \qquad \qquad\qquad \qquad \qquad \qquad \qquad \qquad \ \ \ \ \ \ (IR)
\end{eqnarray*}
for all $(\theta,a)$ satisfying $CE(0,0;\theta,a)=s$ and all  $s\in [\underline{s},\overline{s}]$.
The schedule $[t(s),dd(s)]$  can be converted into the nonlinear premium $t_+(dd)\equiv t[s^{-1}(dd)]$ which is decreasing and convex in deductible, i.e. the marginal price for higher coverage is increasing. This is similar to the concavity of tariff in nonlinear pricing models. See (say) Tirole (1988).

We note that for each $s$ there must exist at least one 
$(\theta,a)$-individual or equivalently $(\theta(s),a(s))$-individual for whom the (IC) constraint binds. Thus, for this individual the (IC) constraint can be written   as
\begin{eqnarray*}
\max_{\tilde{s}\in[\underline{s},\overline{s}]}\!CE(t(\tilde{s}), dd(\tilde{s});\!\theta(s),\!a(s))\!\!=\!\!\max_{\tilde{s}\in[\underline{s},\overline{s}]}\!w\!-\!
t(\tilde{s})\!-\!\frac{\theta(s)\! \left[\int_{0}^{dd(\tilde{s})}\!e^{a(s)D} dH(D)\!+\!e^{a(s)dd(\tilde{s})}[1\!-\!H(dd(\tilde{s}))]\!-\!1\right]}{a(s)},
\end{eqnarray*}
leading to the local (IC) constraint given by the  first-order condition at $\tilde{s}=s$
\begin{eqnarray*}
t'(s) = - \theta(s) e^{a(s) dd(s)} [1-H(dd(s))] dd'(s),
\end{eqnarray*}
where $\theta(s)= a(s)(w-s)/[\phi_a-1]$ and the prime indicates a derivative. This gives
\begin{eqnarray}
dd'(s) = -\eta(s,a(s),dd(s))  t'(s),
\end{eqnarray}
for all $s\in [\underline{s},\overline{s}]$, where   
\begin{eqnarray}
\eta(s,a(s),dd(s))= \frac{\phi_a-1}{a(s)(w-s) e^{a(s)dd(s)} [1-H(dd(s))]}>0.
\end{eqnarray}
Equation (8) is the local incentive compatibility constraint for the insurer's optimization problem.
Regarding the individual rationality constraint, in view of the previous discussion, the  $\overline{s}$-individual has the largest  outside option of no insurance.  Thus, the insurer should bind the (IR) constraint for this individual and make him indifferent between buying insurance or not. This gives  the (IR) constraint 
\begin{eqnarray}
CE(t(\overline{s}),dd(\overline{s});\underline{\theta},\underline{a})=\overline{s}.
\end{eqnarray}

We can now solve the insurer's problem which is to maximize his expected 
profit (7)  subject to  (8) and (10).  Applying the Pontryagin principle (see Appendix), 
the optimal coverage  $(t(s),dd(s))$ is solution of
\begin{eqnarray}
&&  \eta(s, a(s),dd(s)) {\rm E}(\theta|s)[1-H(dd(s))] \nonumber
\\&& \quad +\frac{K(s)}{k(s)}\frac{1}{\eta(s,a(s),dd(s))}
\left[- \frac{\partial \eta(s,a(s),dd(s))}{\partial dd}  dd'(s) +\eta'(s,a(s),dd(s))\right]\!=\!1,\\
&&dd'(s)= -\eta(s, a(s),dd(s))  t'(s),
\end{eqnarray}
where $\eta'(s, a(s),dd(s))$ denotes the total derivative of $\eta(s,a(s),dd(s))$ with respect to $s$,
with the boundary condition $CE(t(dd(\overline{s})), dd(\overline{s}); \underline{\theta},\underline{a})= \overline{s}$.
Evaluating (11) at $\underline{s}$, i.e. for the $(\overline{\theta},\overline{a})$ individual, shows that $dd(\underline{s})=0$, i.e. the highest risk/risk averse individual is offered full coverage as in the benchmark model of Stiglitz (1977).\footnote{Because $K(\underline{s})=0$, (11) and (9) give $[\phi_{\overline{a}}-1] {\rm E}(\theta|\underline{s})/[\overline{a}(w-\underline{s}) e^{\overline{a}dd(\underline{s})}]=1$. Using (5) and ${\rm E}(\theta|\underline{s})=\overline{\theta}$ give $e^{\overline{a}dd(\underline{s})}=1$.}
The next lemma implies that the deductible at equilibrium is increasing in $s$.

\medskip\noindent
{\bf Lemma 2:} {\em  An insurance contract $[t(s),dd(s)]$ satisfies the (IC) constraint if and only if $dd(s)$ is increasing in $s$.}

\medskip\noindent
Since the equilibrium contract satisfies the (IC) constraint, its deductible is increasing in $s$. In other words, individuals with lower risk and/or risk aversion have lower coverage with a larger deductible.

\medskip\noindent
{\sc Finite Number of Contracts}

The principal may offer a finite number $C$ of contracts from which the agent can choose. To simplify the presentation, we consider  $C=2$, where $C$ is exogenous.
Let $(t_1,dd_1)$ and $(t_2,dd_2)$ with $t_1< t_2$ and $dd_1>dd_2$ be these two contracts. We show how the insurer can determine 
these two contracts optimally. 
In addition to the pooling of pairs $(\theta,a)$ leading to the same certainty equivalence $s$, there is bunching of agents with different values of $s$. 
 
The insurer chooses $(t_1,dd_1,t_2,dd_2)$ to maximize his expected profit.
Let ${\cal S}_c$ be the set of agents choosing the contract $(t_c,dd_c)$, $c=1,2$.
Similarly to (7), we have 
\begin{eqnarray*}
{\rm E}[\pi] = \sum_{c=1}^2 \int_{{\cal S}_c} \left[ t_c-  \theta  \int_{dd_c}^{\overline{d}}\! (1\!-\!H(D)) dD   \right]\!dF(\theta,a) 
   =    \sum_{c=1}^{2}\nu_c   \left[t_c-{\rm E}[\theta|{\cal S}_c] \int_{dd_c}^{\overline{d}} \!(1\!-\!H(D)) dD     \right]\!,
\end{eqnarray*}
where the second equality follows from  $\int_{{\cal S}_c} \theta dF(\theta,a)= \nu_c {\rm E}[\theta|{\cal S}_c]$ with $\nu_c= \int_{{\cal S}_c} dF(\theta,a)$ being
the proportion of insurees choosing contract $c$.
The optimal contracts  also need to  satisfy the  incentive compatibility and participation   constraints:
\begin{eqnarray*}
&& CE(t_{c},dd_{c};\theta ,a)\geq CE(t_{c'},dd_{c'},\theta,a),  c\neq c',  \quad \forall (\theta,a)\in {\cal S}_c, c=1,2, \\
&& CE(t_{c},dd_{c};\theta, a)\geq CE(0,0;\theta ,a), \quad \forall (\theta ,a)\in {\cal S}_c, c=1,2.
\end{eqnarray*}

The (IC) constraint reduces to two subsets ${\cal S}_1$ and ${\cal S}_2$ that partition $\Theta \times {\cal A}$ such that  
individuals in  ${\cal S}_1$  and ${\cal S}_2$ choose $(t_1,dd_1)$ and $(t_2,dd_2)$, respectively.
The frontier between ${\cal S}_1$ and ${\cal S}_2$ is determined by the locus of $(\theta,a)$-insurees who are indifferent between the two contracts,
i.e., for whom $CE(t_1,dd_1;\theta,a)= CE(t_2,dd_2;\theta,a)$. 
Using (6), the frontier is  the strictly decreasing  curve in $\Theta \times {\cal A}$ defined by   
\begin{eqnarray} 
\theta(a)&=& \frac{a(t_2-t_1)}{\left[\int_{0}^{dd_1}e^{aD}dH(D)+e^{a dd_1}(1-H(dd_1))-\int_{0}^{dd_2}e^{aD}dH(D)-e^{a dd_2}(1-H(dd_2))\right]}\nonumber\\
   &=& \frac{t_2-t_1}{ \int_{dd_2}^{dd_1} e^{aD} (1-H(D)) dD},
 \end{eqnarray}
where the second equality uses integration by parts.
Regarding the (IR) constraints,  the only one that binds is for the $(\underline{\theta},\underline{a})$-insuree, i.e.  $CE(t_1,dd_1;\underline{\theta},\underline{a}) =  \overline{s}$.

Maximizing ${\rm E}[\pi]$ with respect to $(t_1,dd_1,t_2,dd_2)$
subject to the (IC) and (IR) constraints gives the   first-order conditions 
\begin{eqnarray}
&&\!\!\!\!\!\!\nu_1 +\int_{\underline{a}}^{a^*}\!\left[t_{1}\!-\!\theta(a)\left\{\int_{dd_1}^{\overline{d}}(1\!-\!H(D)) dD\right\}\right]f(\theta(a),a)\frac{\partial \theta(a)}{\partial t_1} da  \nonumber \\
&& \!\!\!\!\!\quad -\int_{a^*}^{\overline{a}}\!\left[t_2\!-\!\theta(a)\left\{\int_{dd_2}^{\overline{d}}(1\!-\!H(D)) dD \right\}\right]f(\theta(a),a)\frac{\partial \theta(a)}{\partial t_1} d a=\rho\\
&&\!\!\!\!\!\!\int_{\underline{a}}^{a^*}\!\left[t_1\!-\!\theta(a)\left\{\int_{dd_1}^{\overline{d}}(1\!-\!H(D)) dD\right\}\right]\!\!f(\theta(a),a)\frac{\partial \theta(a)}{\partial dd_1} da+{\rm E}[\theta|{\cal S}_1]\nu_1 (1\!-\!H(dd_1)) \nonumber\\
&&\!\!\!\!\! -\!\!\int_{{a}^*}^{\overline{a}}\!\left[t_2\!-\!\theta(a)\!\left\{\!\!\int_{dd_2}^{\overline{d}}(1\!-\!H(D)) dD\right\}\right]\!\!f(\theta(a),a)
\frac{\partial \theta(a)}{\partial dd_1} da   -\!\rho\underline{\theta}e^{\underline{a}dd_1}(1\!-\!H(dd_1))=0\\
&&\!\!\!\!\!\!\int_{\underline{a}}^{a^*}\!\left[t_1\!-\!\theta(a)\left\{\int_{dd_{1}}^{\overline{d}}(1\!-\!H(D)) dD\right\}\right]f(\theta(a),a)\frac{\partial \theta(a)}{\partial t_2} da   \nonumber \\
&&\!\!\!\!\!+\nu_2-\int_{a^*}^{\overline{a}}\!\left[t_2\!-\!\theta(a)\left\{\int_{dd_{2}}^{\overline{d}}\!(1-H(D))dD\right\}\right]f(\theta(a),a)\frac{\partial \theta(a)}{\partial t_2}da=0\\
&&\!\!\!\!\!\!\int_{\underline{a}}^{a^*}\!\left[t_1\!-\!\theta(a)\left\{\int_{dd_{1}}^{\overline{d}}(1\!-\!H(D)) dD\right\}\right]f(\theta(a),a)
\frac{\partial \theta(a)}{\partial dd_2}da+{\rm E}(\theta|{\cal S}_2)\nu_2(1\!-\!H(dd_2))  \nonumber\\
&& \!\!\!\!\!-\int_{a^*}^{\overline{a}}\!\left[t_2\!-\!\theta(a)\left\{\int_{dd_{2}}^{\overline{d}}(1\!-\!H(D)) dD\right\}\right]f(\theta(a),a)\frac{\partial \theta(a)}{\partial dd_2} da=0,\\
&&\!\!\!\!\!\!t_1=\frac{\underline{\theta}}{\underline{a}}\!\left[\int_{dd_1}^{\overline{d}}\left(e^{\underline{a}D}-e^{\underline{a}dd_1}\right)dH(D)\right]
\end{eqnarray}
where  $\rho$ is  the Lagrangian multiplier associated with the (IR) constraint and $a^*$ is  the  minimum of $\overline{a}$ and the value which solves  (13) evaluated at $\underline{\theta}$.

\medskip\noindent
{\sc Extensions}

Our model extends to other insurance contracts such as health.
Up to some variations, health insurance  involves a premium $t$ as well as a per period deductible $dd$ and a copayment $\gamma$ per medical procedure/visit. In particular, and in contrast to the automobile insurance, the deductible is not per visit while the copayment arises in the first procedure/visit after the deductible is met. 
In this case,  for a contract $[t(\theta,a),dd(\theta,a),\gamma(\theta,a)]$, the insurer's expected profit (1) becomes
\begin{eqnarray*}
{\rm E}[\pi(\theta,a)]&=& \int_{\Theta \times {\cal A}} \left\{ t(\theta,a) -
{\rm E}\left[\Unit(D_1+\ldots+D_J > dd(\theta,a)) \left(D_1+\ldots+D_J
\right. \right. \right.\\
&& \qquad\ \qquad \quad \left. \left. \left. -dd(\theta,a)- \gamma(\theta,a) (J-J^{\dagger})   \right)      \right]   \right\} dF(\theta,a),
\end{eqnarray*}
where the expectation in the integral is with respect to the total expense $D_1+\ldots+D_J$, the number $J$ of visits and $J^{\dagger}$ which is 
the minimal number of visits for which the deductible is met, i.e., 
$J^{\dagger} = {\rm argmin}_{j=1,\ldots,J} D_1+\ldots+D_j > dd$.  The per visit expenses $D_j, j=1,\ldots,J$ may no longer be independent.
Indeed, a patient with a medical condition will exhibit correlated medical expenses  over the treatment period.  Similarly, the per visit expense $D_j$ might be correlated with the expected number of medical procedures/visits $\theta$.

Regarding the patient,  his expected utility (2) without health insurance 
becomes
\begin{eqnarray*}
V(0,0,0;\theta,a)=  {\rm E}[U(w-D_1-\ldots-D_J;a)]= -e^{-aw} {\rm E}[e^{-a(D_1+\ldots+D_J)}],
\end{eqnarray*}
under the CARA utility function by A1-(i),
where the expectation is with respect to the total expense $D_1+\ldots+D_J$
and the number $J$ of visits which depends on $\theta$.
The expected utility (3) of a $(\theta,a)$-patient buying coverage $(t,dd,\gamma)$ becomes 
\begin{eqnarray*}
V(t,dd,\gamma;\theta,a) = - e^{-aw} {\rm E}[e^{-a X}],
\end{eqnarray*}
where $X$ is the out-of-pocket expense $X= (D_1+\ldots+D_J) \Unit(D_1+\ldots+D_J \leq dd) + (dd + (J-J^{\dagger}) \gamma)\Unit(D_1+\ldots+D_J > dd)$.
When there is a finite number $C$ of offered coverages, the insurer partitions the set of types $\Theta \times {\cal A}$ using the patients' certainty equivalences to maximize his expected profit  with respect to the contract terms $(t_c,dd_c,\gamma_c), c=1,\ldots,C$.

\section{Identification with a Continuum of Contracts}

In this section, we consider  the  case in which a continuum of coverages is offered to each insuree. In particular, our identification analysis shows the key role played by the number of accidents.
The model structure is given by the joint distribution of risk and risk aversion $F(\cdot,\cdot)$ and the damage distribution $H(\cdot)$. Besides the specification of the CARA utility function  and the Poisson distribution for the number of accidents, the identification problem is nonparametric.\footnote{The problem of identifying nonparametrically the agent's utility function is quite complex. 
In the context of auctions, the bidder's utility function is not identified in general. Nonparametric identification is achieved  with the help of exclusion restrictions using exogenous variations in the number of bidders as in Guerre, Perrigne and Vuong (2009) or with the help of additional data from ascending auctions  as in Lu and Perrigne (2008).
See also Campo, Guerre, Perrigne and Vuong (2009) for semiparametric identification when the bidder's utility function is parameterized as CARA or CRRA.}
The problem of identification is to recover uniquely the structure $[F(\cdot,\cdot),H(\cdot)]$ from the observables. In the case of  a continuum  of contracts, we observe  the contract purchased by each insuree $(t,dd)$ and  the $J$ claims made by each insuree with the corresponding amounts of damages
$(D_1,\ldots,D_J)$. In Section 3.2, we observe $J^*$ claims with their corresponding damages $(D_1,\ldots,D_{J^*})$ because of the truncation 
at the deductible.

We introduce some  observed variables $X$ characterizing the insuree and his/her car that are used by the insurer to discriminate insurees.\footnote{Variables that are not used to discriminate insurees can enter in the model through $(\theta,a)$ which can be then viewed as aggregating  observed and unobserved heterogeneity.} Variables related to the insuree 
may contain age, gender, education, marital status, location and driving experience.  Variables 
related to the insuree's  car may include  car mileage, business use, car value, power, model and make.\footnote{We can use the car value as a proxy for wealth $w$  so that $w$ is a variable in $X$.} 
With the introduction of $X$ with values in the support  ${\cal S}_{X} \subset \Real^{\dim X}$, the model structure becomes $[F(\theta,a|X), H(D|X)]$ as we expect that such variables  affect 
the insuree's risk and risk aversion as well as the damage.
For instance, the damage  with  an  expensive car is  likely to be larger than the damage with an inexpensive one.  
Let $G(\cdot|X)$ denote the observed deductible distribution conditional on $X$. It is crucial that all the variables  used by the insurer to discriminate insurees are included in $X$.

In identification studies  of structural models, it is important to define the set of admissible structures that are consistent with the assumptions of the theoretical model.
We formalize such assumptions on the structure and $(\theta,a,J,D,X)$. 
Specifically, the structure $[F(\cdot,\cdot|X),H(\cdot|X)]$ belongs to  ${\cal F}_{X}\times {\cal H}_{X}$ as defined below.

\medskip\noindent
{\bf Definition 1:} {\em Let ${\cal F}_{X}$ be the set of conditional distributions $F(\cdot,\cdot|X)$ satisfying\\
(i) For every $x \in {\cal S}_{X}$, $F(\cdot,\cdot|x)$ is a c.d.f. with compact support $\Theta(x) \times {\cal A}(x)= [\underline{\theta}(x),\overline{\theta}(x)] \times [\underline{a}(x),\overline{a}(x)] \subset\Real_{++}\times\Real_{++}$, \\
(ii) The conditional density  $f(\cdot,\cdot|\cdot) >0$ on its support.
}

\medskip\noindent
{\bf Definition 2:} {\em Let ${\cal H}_{X}$ be the set of distributions $H(\cdot|X)$ satisfying\\
(i) For every $x \in {\cal S}_{X}$, $H(\cdot|x)$ is a c.d.f with compact  support  $[0,\overline{d}(x)] \subset\Real_+$  with $\sup_{x\in {\cal S}_{X}} \overline{d}(x)<+\infty$, \\
(ii)  The conditional density $h(\cdot|\cdot) > 0$ on its support.
}

\medskip\noindent
{\bf Assumption A3:} {\em We have\\
(i) $(D_1,\ldots,D_J) \perp (\theta,a)  \big{|}(J,X)$.\\
(ii) $ (D_1,\ldots,D_J) \big{|} (J,X)$ are i.i.d. as $H(\cdot|X)$,\\
(iii)  $J \perp (X,a)\big{|} \theta$ with $J|\theta \sim {\cal P}(\theta)$, i.e. ${\rm Pr}[J=j]=e^{-\theta}\frac{\theta^j}{j!}$,\\
(iv) $(\theta,a,J,X)$ is i.i.d. with $(\theta,a)|X \sim F(\cdot,\cdot|X)$
}

\medskip\noindent
Assumption A3 parallels A1 with $X$.
Assumption A3-(i) implies that  conditional on  $X$, the  amount of damage does not provide any information on his risk and risk aversion.   For instance, conditional on $X$, damages depend on exogenous factors that are independent of $(\theta,a)$. In the same spirit, Assumption A3-(ii) says that damages are mutually independent conditional  on $X$. 
Regarding Assumption A3-(iii), the number of accidents  $J$ depends on the insuree's risk $\theta$ only, while the Poisson distribution follows the theoretical 
model of Section 2, where the insuree's risk $\theta$ is the expected number of accidents.  By Assumption A3-(iv), $(\theta,a,J,X)$  is i.i.d. across insurees.
We maintain Assumption A3 throughout the paper. 
Lastly, this section assumes that the observed $(t,dd)$ correspond to the  optimal coverage schedule  so that (11) and (12) are satisfied.

\subsection{Case 1: Full Damage Distribution}

Case 1 considers  a continuum of coverages offered to each insuree as well  the observation of damage for every accident whether  it is below or above the deductible. 
It follows that $H(\cdot|X)$ is identified  on $[0,\overline{d}(X)]$. 
It remains to study the identification of $F(\cdot,\cdot|X)$. For the rest of Section 3, to simplify the notations, we suppress the conditioning on $X$.
We first proceed by studying the identification of the distribution $K(\cdot)$ of certainty equivalence (5) of no coverage.
The optimal contracts are characterized by  (11) and (12). Equation (11) defines a one-to-one mapping between the certainty equivalence $s$
and the deductible $dd$, while (12) defines a one-to-one mapping between $dd$ and $t$. 
The key idea is to exploit
the former  mapping to identify the distribution of certainty equivalence from the observed deductible distribution $G(\cdot)$.
This result is in the spirit of the nonparametric identification literature on auctions and contracts.\footnote{For auctions, see 
Guerre, Perrigne and Vuong (2000) and  Athey and Haile (2007) where the mapping between the observed bid and the unobserved private value identifies the private value distribution. For contracts, see Luo, Perrigne and Vuong (2015) in the context of nonlinear pricing,  and Perrigne and Vuong (2011) in the context of a procurement model with adverse selection and moral hazard.  The mapping between the observed quantity/ price and the unobserved consumer's type/firm's efficiency is exploited to recover their underlying distribution, respectively.}
We have $G(dd)= {\rm Pr}(\widetilde{dd}\leq dd)={\rm Pr}(s(\tilde{dd})\leq s(dd))={\rm Pr}(\tilde{s}\leq s(dd))=K(s)$ implying
$g(dd)=k(s)s'(dd)$, with $s(\cdot)$ being the inverse of $dd(\cdot)$  by monotonicity of the latter.
Hence,
\begin{eqnarray*}
\frac{G(dd)}{g(dd)}=\frac{K(s)}{k(s)}\frac{1}{s'(dd)}=  \frac{ K(s)}{k(s)} dd'(s).
\end{eqnarray*}

Substituting the above expression in (11), we obtain 
\begin{eqnarray*}
\eta(s, a(s),dd(s)){\rm E}[\theta|s](1\!-\!H(dd)) 
\!+\!\frac{G(dd)}{g(dd)} \left\{\! -\frac{\frac{\partial \eta(s,a(s),dd(s))}{\partial dd}}{\eta(s, a(s),dd(s))}
\!+\! \frac{\eta'(s,a(s),dd(s))}{\eta(s,a(s),dd(s))}s'(dd)\! \right\}\!=\!1.
\end{eqnarray*}
From (12), we have $t'_+(dd)\!=\!-1/\eta(s, a(s),dd(s))$, where $t_+(dd)\equiv t[s^{-1}(dd)]$ is the function relating the deductible to the premium. We also have
$d t'_+(dd(s))/ds=-d [\eta(s, a(s),dd(s))]^{-1}$ $/ds$, i.e. $t_+^{\prime\prime}(dd) \times dd'(s)= \eta'(s, a(s),dd(s))/[\eta(s, $  $a(s),dd(s))]^2$ or equivalently
$t_+^{\prime\prime}(dd)=[\eta'(s, a(s),dd(s))/[\eta(s, a(s),dd(s))]^{2}] \times s^{-1'}(dd).$
Using this result, we can rewrite the previous equation as
\begin{eqnarray*}
{\rm E}[\theta|s](1-H(dd))+\frac{G(dd)}{g(dd)} \left\{- \frac{\frac{\partial \eta(s,a(s),dd(s))}{\partial dd}}{\eta(s,a(s),dd(s))^2}+  t_+^{\prime\prime}(dd) \right\} = -t'_+(dd).
\end{eqnarray*}
From the definition (9) of $\eta(\cdot,\cdot,\cdot)$, its partial derivative  with respect to $dd$ is
\begin{eqnarray*}
\frac{\partial \eta(s,a(s),dd(s))}{\partial dd}=-\eta(s,a(s),dd(s))\left[a(s)-\frac{h(dd)}{1-H(dd)}\right].
\end{eqnarray*}
Thus, the first-order condition defining the optimal deductible can be rewritten as
\begin{eqnarray*}
{\rm E}[\theta|dd](1-H(dd))+\frac{G(dd)}{g(dd)}\left[ -t'_+(dd) \left(a(s) - \frac{h(dd)}{1-H(dd)} \right)+ t_+^{\prime\prime}(dd) \right] =-
t'_+(dd),
\end{eqnarray*}
where ${\rm E}[\theta|s]={\rm E}[\theta|dd]$ because of the one-to-one mapping between $dd$ and $s$. 
After elementary algebra, we obtain
\begin{eqnarray*}
a(s)= \frac{1}{t'_+(dd)}    \left\{ 
            \frac{g(dd)}{G(dd)} 
           \left[  t'_+(dd) +  {\rm E}[\theta|dd] (1-H(dd)) \right] 
             + t_+^{\prime\prime}(dd)          \right\}                   +  \frac{h(dd)}{1-H(dd)},
\end{eqnarray*}
showing that $a(s)$ is identified as the right-hand side is observed or identified from observables. In particular, ${\rm E}[\theta|dd]$
is identified by the expected number of claims made by insurees choosing the deductible $dd$ given that all the claims are observed, i.e. ${\rm E}[\theta|dd]={\rm E}[J|dd]$.\footnote{We have ${\rm E}[J|dd]={\rm E}[J|s]= 
{\rm E}\{{\rm E}[J|\theta,s]|s \}= {\rm E}\{{\rm E}[J|\theta,a]|s\}= {\rm E}\{{\rm E}[J|\theta]|s \}={\rm E}[\theta|s]$, where we have used A3-(iii) and the one-to-one mapping between $(\theta,a)$ and $(\theta,s)$.}  Then, using (8) and  (9) we have
\begin{eqnarray*}
s= w + \frac{t'_+(dd)(\phi_a -1)}{ a(s) \exp(a(s)dd)  (1-H(dd))},
\end{eqnarray*}
showing that the insuree's certainty equivalence  $s$  can be identified from his choice of deductible  $dd$ and the knowledge of $H(\cdot)$, $G(\cdot)$, $t_+(\cdot)$ and ${\rm E}[J|dd]$. Thus, we have the following result.
 
\medskip\noindent
{\bf Lemma 3:} {\em Suppose that a continuum of optimal insurance coverages is offered to each insuree and  all accidents are observed. Under A3, the pair $[K(\cdot),H(\cdot)]$ is identified.}

\medskip
It remains to investigate whether we can identify $F(\cdot,\cdot)$. A sketch of the argument is as follows.  From the moment generating function of the number of accidents $J$ conditional on $s$, 
we  identify the moment generating function of $\theta$ given $s$ in a neighborhood of zero. As is well known, the latter identifies 
$F_{\theta|S}(\cdot|\cdot)$.   Once we identify $F_{\theta|S}(\cdot|\cdot)$, we use $K(\cdot)$ to  derive the joint distribution of $(\theta, s)$. Identification of the joint density of $(\theta,a)$ follows from the known one-to-one mapping between $(\theta,s)$ and $(\theta,a)$ given by (5).  It is important to note that the observed number of claims $J$ plays a crucial role in identifying $F_{\theta|S}(\cdot|\cdot)$. This is possible because the Poisson distribution  belongs to the class of distributions
whose nonparametric mixture is identified. See Rao (1992). 
In contrast, if one only observes   whether there is an accident
with the risk   measured by the probability of such contingency $\tilde{\theta}=1-e^{-\theta}$, then $F_{\theta|S}(\cdot|\cdot)$  is not identified because the nonparametric mixture of a Binomial distribution does not belong to the aforementioned class and thus  is not identified.  See Aryal, Perrigne and Vuong (2009).

Formally, for a given  certainty equivalence $s$, the subpopulation of insurees with  coverage $(t(s), dd(s))$ and their corresponding claims give the moment generating function $M_{J|S}(\cdot|s)$ as
\begin{eqnarray}
M_{J|S}(t|s) &=& {\rm E}[e^{Jt}|S=s] = {\rm E}\left\{{\rm E}[e^{Jt}|\theta,S]|S=s \right\} \nonumber\\
&=& {\rm E}\left\{{\rm E}[e^{Jt}|\theta,a]|S=s \right\}= {\rm E}\left\{{\rm E}[e^{Jt}|\theta]|S=s \right\} \nonumber\\
&=&  {\rm E} \left\{e^{\theta(e^{t}-1)}|S=s   \right\} 
= M_{\theta|S} (e^{t}-1 | s) ,   
\end{eqnarray}
where the third equality follows from the one-to-one mapping between $(\theta,s)$ and $(\theta,a)$ and  the fourth  and fifth equalities from A3-(iii) using the moment generating function of the Poisson distribution with parameter $\theta$. In particular, (19) shows that the moment generating function $M_{J|S}(\cdot|s)$ exists for every $t \in \Real$ because
$\theta$ has a compact support given $S=s$. Moreover, letting $u=e^t-1$ shows that 
\begin{eqnarray*}
M_{\theta|S}(u|s)= M_{J|S}(\log(1+u)|s)
\end{eqnarray*}
for all $u \in (-1,+\infty)$. Thus $M_{\theta|S}(\cdot|s)$ is identified on a neighborhood of 0 thereby identifying $F_{\theta|S}(\cdot|s)$. See e.g.
Billingsley (1995, p. 390).\footnote{Alternatively, because $M_{\theta|S}(\cdot|s)$ exists in a neighborhood of 0, then all the moments of $\theta$ given 
$S=s$ are identified by $M^{(k)}_{\theta|S}(0|s)={\rm E}[\theta^k|S=s]$ for  $k=0,1,\ldots,\infty$. Since $\theta$ given $s$ has compact support, we are in the class of Hausdorff moment problems, which are always determinate, i.e., the distribution of $\theta$ given $s$ is uniquely determined by its moments.}


The joint density of $(\theta,s)$ is $f(\theta,s)=f(\theta|s) k(s)$, which is identified.
From the  known one-to-one mapping  $T(\cdot,\cdot)$ that transforms $(\theta,a)'$ into $(\theta,s)'$, namely 
$T(\theta,a)=[\theta, w-[\theta (\phi_a-1)]/a]'$ with $\phi_a=\int e^{aD}dH(D)$ and $H(\cdot)$ known,
we  recover $f(\theta,a)$ as
\begin{eqnarray*}
f(\theta,a)=f_{\theta S}(T^{-1}(\theta, a))\Bigg|\frac{\partial T^{-1}(\theta,a)}{\partial (\theta, a)}\Bigg|.
\end{eqnarray*}
This result is formally stated in the following proposition.

\medskip\noindent 
{\bf Proposition  1:} {\em Suppose that a continuum of optimal insurance coverages is offered to each insuree and  all  accidents are observed. Under A3, the structure $[F(\cdot,\cdot),H(\cdot)]$ is identified.}

\subsection{Case 2: Truncated Damage Distribution}

We maintain the assumption that the insurer offers a continuum of  optimal contracts to each insuree but we now consider that the damage distribution is not fully observed.
Making abstraction of dynamic considerations, an accident leads to a claim if and only if the damage is above the  deductible.
Thus, we can identify the  truncated damage distribution on $[dd,\overline{d}]$. However, the deductible $dd$ varies across insurees.
In particular, for insurees buying full insurance, the deductible is zero thereby identifying the damage distribution on its  full support $[0,\overline{d}]$.  Formally, $H_{D|dd}(\cdot|0)= H_{D|S}(\cdot|\underline{s})=
H_{D|(\theta,a)}(\cdot|\overline{\theta},\overline{a})=H_D(\cdot)$ by A3-(i). Thus, we have the following lemma.

\medskip\noindent
{\bf Lemma 4:} {\em Under A3, $H(\cdot)$ is  identified.}

\medskip 
It remains to study the identification of $F(\cdot,\cdot)$. Though 
the reported number of accidents $J^*$ is observed, instead of the true $J$, the argument is similar to Case 1. Specifically,
reviewing the argument leading to Lemma 3,  $K(\cdot)$ is identified if ${\rm E}[\theta|dd]$ is. 
Since accidents are reported only if the damage is above the deductible,  we  have ${\rm E}[\theta|dd] \neq {\rm E}[J^*|dd]$, where 
$J^*$ is the number of reported accidents.
But $J^*$ given $(J,dd)$ is  distributed as a Binomial with parameters  $(J,1-H(dd))$ by A3-(i,ii).
Thus, ${\rm E}[J^*|dd]= {\rm E} \{{\rm E}[J^*|J,dd]|dd\}= {\rm E}[J(1-H(dd))|dd]= (1-H(dd)) {\rm E}[J|dd]= (1-H(dd)){\rm E}[\theta|dd]$, 
i.e. ${\rm E}[\theta|dd]= {\rm E}[J^*|dd]/(1-H(dd))$.
Hence, ${\rm E}[\theta|dd]$ is identified despite the truncated  damage distribution  leading to the identification of $K(\cdot)$.

Turning to the identification of $F(\theta,a)$, we proceed as in Section 3.1. The moment generating 
function of $J^*$ given $s$ is 
\begin{eqnarray}
M_{J^*|S}(t|s) &=& {\rm E}[e^{J^*t}|S=s] = {\rm E}\{{\rm E}[e^{J^*t}|J,S]|S=s \} = {\rm E}\{{\rm E}[e^{J^*t}|J,dd]|S=s \} \nonumber \\
&=&  {\rm E} \left\{[H(dd) + (1-H(dd))e^{t}]^J|S=s \right\} = {\rm E}\left\{ e^{J \log [H(dd)+(1-H(dd))e^t]}|S=s    \right\} \nonumber\\
&=&  M_{\theta|S} \left[e^{\log [H(dd) + (1-H(dd))e^t]} -1|s   \right]  = M_{\theta|S} [(1-H(dd))(e^t-1)|s],
\end{eqnarray}
where the fourth equality uses the moment generating function of the Binomial distribution ${\cal B}(J,1-H(dd))$, and the fifth equality 
uses (19) with $t$ replaced by $\log [H(dd)+(1-H(dd))e^t]$.
Thus, we obtain
\begin{eqnarray*}
M_{\theta|S}(u|s) = M_{J^*|S} \left[\log \left(1 + \frac{u}{1-H(dd)}   \right)\Big{|}s    \right],
\end{eqnarray*}
for $u\in (-(1-H(dd),+\infty)$. The rest of the argument in Case 1 applies leading to the following proposition.

\medskip\noindent 
{\bf Proposition  2:} {\em Suppose that a continuum of  optimal insurance coverages is offered to each insuree and  accidents are observed if and only if the damage is above the deductible. Under A3, the structure $[F(\cdot,\cdot),H(\cdot)]$ is identified.}

\section{Identification with a Finite Number of Contracts}

We now address the identification of the model when only (say) two contracts are offered given $X$. The identification argument can no longer rely on the identification of the density of certainty equivalence as we cannot exploit the one-to-one mapping between the insuree's certainty equivalence and his deductible choice. There is a continuum of $s \in [\underline{s},\overline{s}]$ values, while there are only a finite number of deductibles.
Consequently,  the FOCs (14)--(18) characterizing  $(t_1,dd_1,t_2,dd_2)$  will not allow us to identify $F(\theta,a)$.
In addition to the key role played by the observed number of claims, 
we  exploit sufficient variations in exogenous variables to achieve identification.
A notable feature of Section 4 is that we do not require that the  observed coverages $(t_1,dd_1,t_2,dd_2)$ are optimal. Consequently, the results of this section apply beyond the case of monopoly  to entertain data from other forms of competition among insurers.
As before, we distinguish whether the full  or truncated damage distribution is observed. Regarding observables, for each insuree we need  the  pair of offered coverages  $(t_1,dd_1,t_2,dd_2)$,  his choice of coverage, the number of accidents, their corresponding damages and the characteristics $X$.

\subsection{Case 3: Full Damage Distribution}

This case is the closest to  Cohen and Einav (2007) who  identify the joint distribution of risk and risk aversion under parametric assumptions.  In this section, we show how  insuree's optimal coverage choice  with a full support 
assumption  and sufficient variations in some exogenous characteristics 
can identify nonparametrically $f(\theta,a)$. In view of Cohen and Einav (2007) empirical findings,  our identification result is important for several reasons. 
First, the nonparametric identification of the joint distribution of risk and risk aversion  offers more flexibility on the dependence between 
risk and risk aversion. Their empirical findings display a counterintuitive positive correlation between the latter. Second, their robustness analysis suggests that the offered contracts are suboptimal with their estimated positive correlation, i.e., the insurer 
could increase his profit by adjusting upward the current low deductibles that are more compatible with a negative correlation. 

Our identification results rely   on a nonparametric mixture of a Poisson distribution for the number of claims. Specifically, the probability of the observed claims $J$ conditional on the characteristics $x$ is given by
\begin{eqnarray*}
{\rm Pr}[J=j|x]= \int_{\underline{\theta}(x)}^{\overline{\theta}(x)} e^{-\theta} \frac{\theta^j}{j!} dF_{\theta|X}(\theta|x) 
\end{eqnarray*} 
where the mixing distribution $F_{\theta|X}(\cdot|x)$ is left unspecified. 
Given that all the accidents and   damages are observed, the damage distribution $H(\cdot|X)$ is identified.
To establish identification of $F(\theta,a|X)$, we proceed as follows. 
We first show the identification of  $F_{\theta|X}(\cdot|\cdot)$ 
 following an argument similar to Case 1. In  the second step,  we identify the conditional distribution $F_{a|\theta,X}(\cdot|\cdot,\cdot)$  
at the frontier $a(\theta,X)$  between the two sets ${\cal S}_1(X)$ and ${\cal S}_2(X)$ that partition  $\Theta(X) \times {\cal A}(X)$ according to the coverage choices of insurees with characteristics $X$. 
In the third step, we make an exclusion restriction and a full support assumption 
involving  some  characteristics $Z$  included in $X$  to achieve identification of the distribution $F_{a|\theta,X}(\cdot|\cdot,\cdot)$  on its support.

For the first step,  we exploit again  the observed number of accidents. Using an argument similar to that leading to (19)  for the subpopulation of insurees with characteristics $x$, the moment generating function $M_{J|X}(\cdot|x)$ is
\begin{eqnarray*}
M_{J|X}(t|x) &=& {\rm E}[e^{Jt}|X=x] = {\rm E}\left\{{\rm E}[e^{Jt}|\theta,X]|X=x, \right\} \nonumber\\
&=& {\rm E}\left\{{\rm E}[e^{Jt}|\theta]|X=x \right\} 
=  {\rm E} \left\{e^{\theta(e^{t}-1)}|X=x \right\} \nonumber\\
&=& M_{\theta|X} (e^{t}-1 | x),   
\end{eqnarray*}
where the third and fourth equalities follow from A3-(iii). Thus,  $f_{\theta|X}(\cdot|\cdot)$ is identified by its moment generating function 
\begin{eqnarray*}
M_{\theta|X}(u|x)= M_{J|X}(\log(1+u)|x)
\end{eqnarray*}
for all $u \in (-1,+\infty)$. 

In the second step, we consider the probability that an insuree with risk $\theta$ and characteristics $X$ chooses the coverage $(t_1(X),dd_1(X))$ as intuitively this provides information about the insuree's risk aversion $a$.
To do so, we define a discrete variable $\chi$, which takes  values 1 and 2  depending on whether the insuree chooses the coverage $(t_1(X),dd_1(X))$
or  $(t_2(X),dd_2(X))$, i.e., whether the insuree's types $(\theta,a)$ belongs to ${\cal S}_1(X)$ or ${\cal S}_2(X)$, respectively.
Thus, $\chi=1$ is also equivalent to $a\leq a(\theta,X)$, where the latter is the inverse  of the frontier (13), where $(t_1,dd_1,t_2,dd_2)$ and $H(\cdot)$ now depends on $X$. Namely, $a(\theta,X)$ is the inverse of
\begin{eqnarray*} 
\theta(a,X)
   &=& \frac{t_2(X)-t_1(X)}{ \int_{dd_2(X)}^{dd_1(X)} e^{aD} (1-H(D|X)) dD}.
 \end{eqnarray*}
Our identification strategy below  exploits variations of this frontier in $X$.  In particular,  even if the deductible does not vary with $X$ as with US data, the premium and possibly the damage distribution do depend on $X$.

The  probability of interest can then be written as ${\rm Pr}[\chi=1|\theta,X=x]$, which is
\begin{eqnarray*}
F_{a|\theta,X}[a(\theta,x)|\theta,x] = \frac{f_{\theta|\chi,X}(\theta|1,x)\nu_1(x)}{ f_{\theta|X}(\theta|x)},
\end{eqnarray*}
by Bayes' rule, where $\nu_1(x)$ is the proportion of insurees with characteristics $x$ choosing the coverage $(t_1(x),dd_1(x))$.
The latter is identified from the data. Since $f_{\theta|X}(\cdot|\cdot)$ is identified from the first step, it remains to identify 
$f_{\theta|\chi,X}(\cdot|1,x)$.
Applying the same argument as in Step 1 but conditioning on $\chi=1$ as well, we obtain 
\begin{eqnarray*}
M_{J|\chi,X}[t|1,x] &=& {\rm E}[e^{Jt}|\chi\!=\!1,X\!=\!x] = {\rm E}\{{\rm E}[e^{Jt}|\theta,a,X]|\chi\!=\!1,X\!=\!x\} \nonumber\\
&=&  M_{\theta|\chi,X}[e^t-1|1,x],   
\end{eqnarray*}
where the second equality follows from the equivalence between  conditioning on $(\theta,a,\chi)$ and conditioning on $(\theta,a)$, while the third equality 
follows, as before, from A3-(iii).
Thus, $f_{\theta|\chi,X}(\cdot|1,\cdot)$ is identified by its moment generating function 
\begin{eqnarray*}
M_{\theta|\chi,X}(u|1,x)= M_{J|\chi,X}(\log(1+u)|1,x)
\end{eqnarray*}
for all $u \in (-1,+\infty)$.
Hence,  $F_{a|\theta,X}[a(\theta,x)|\theta,x]$ is identified for every $\theta \in [\underline{\theta}(x),$  $\overline{\theta}(x)]$ 
and $x \in {\cal S}_{X}$.

To conduct  policy  counterfactuals
the analyst may need to identify $F(\cdot,\cdot|x)$ on the whole support $\Theta(x) \times {\cal A}(x)$.
This is the purpose of the third step.
To do so, we partition the vector $X$ into $(W,Z)$. Let ${\cal S}_{W}$ denote the support of  $W$ and ${\cal S}_{W_1|w2}$ denote the support of some variable $W_1$ given some variable $W_2=w_2$. 

\medskip\noindent
{\bf Assumption A4:} {\em  We have\\
(i) $a \perp Z \big{|} (\theta,W)$\\
(ii) $\forall (\theta,a,w) \in {\cal S}_{\theta a W}$, there exists $z \in {\cal S}_{Z|\theta w}$ such that 
$a(\theta,w,z)=a$.}
 
\medskip
Assumption A4-(i) is an exclusion restriction, i.e. $Z$ does not affect  risk aversion given risk and other characteristics $W$.  
The variable $Z$ needs to be  continuous and can be the car value, the  reported annual mileage, the driver's experience, etc. 
This gives
\begin{eqnarray*}
F_{a|\theta,W,Z}(a(\theta,w,z)|\theta,w,z) = F_{a|\theta,W}(a(\theta,w,z)|\theta,w), \quad\!\! \forall (\theta,w,z).
\end{eqnarray*}
Because the left-hand side is identified from the second step, sufficient variations in $a(\theta,w,z)$ due to $z$ can identify $F_{a|\theta,W}(\cdot|\theta,w)$.
This is the purpose of A4-(ii), which is a full support assumption. 
Similar assumptions (sometimes called large support assumptions) have been   made  in various contexts. See Matzkin (1992, 1993),  Lewbel (2000), Carneiro, Hansen and Heckman (2003),  Imbens and Newey (2009) and  Berry and Haile (2014) among others.  
In our context, this assumption can be interpreted as follows: For every individual with  characteristics $(\theta,a,W)$, there  exists some characteristics $Z$ such as the  car value  or the mileage for which the insuree is indifferent between the two offered coverages.
The full support  assumption is  sufficient to guarantee identification since
\begin{eqnarray*}
F_{a|\theta,W}(a|\theta,w) = F_{a|\theta,W}[a(\theta,w,z)|\theta,w] = F_{a|\theta,W,Z}[a(\theta,w,z)|\theta,w,z],
\end{eqnarray*}
where the first equality uses the full support assumption and the second equality uses the exclusion restriction.
Note that $a(\cdot,\cdot,\cdot)$ is identified in view of (13). The full support assumption guarantees that for every $a$ on its support, there exists a known value $z$  such that $a=a(\theta,w,z)$.
Identification of $F(\theta,a|w,z)$ follows using the first step. This result is formally stated in the next proposition.

\medskip\noindent
{\bf Proposition 3:} {\em   Suppose that two insurance coverages are offered to each insuree and  all accidents are observed for each insuree. Under A3 and A4, the structure $[F(\cdot,\cdot|X),H(\cdot|X)]$ is identified.}

\medskip\noindent
Despite pooling due to both multidimensional screening, and a finite number of coverage, Proposition 3 shows that the model primitives are identified by exploiting wisely  the number of accidents and  variations in some exogenous variable. In particular, our identification argument does not require optimality of the offered coverages. This  is novel  in the identification of models under incomplete information.

\subsection{Case 4: Truncated Damage Distribution}

The data scenario analyzed in Case 4 corresponds to typical insurance data, i.e., a finite number of contracts offered with claims filed only if damages are above the deductible. 
Case 3 has shown that observing a finite number of contracts does not prevent the nonparametric identification of the joint distribution of risk and risk aversion provided   all accident information is available  and  there is enough variation in some excluded exogenous variable. 
In contrast, the truncation on the damage distribution in Case 4  limits the extent of  identification.
Nevertheless, we show that $F(\cdot,\cdot|X)$ is identified up to  the knowledge  of the probability to have a damage below the  lowest deductible, i.e., $H(dd_2(X)|X)$.\footnote{When two contracts are offered, it is never optimal for the insurer to offer full insurance, i.e. $dd_2(X)=0$. Therefore, we cannot use the argument of Case 
2 to identify $H(\cdot|X)$ and hence $H(dd_2(X)|X)$.} To simplify the notations, we let $H_c(X)\equiv H(dd_c(X) |X)$ hereafter.

We note  the relationship between $1-H_1(X)$ and $1-H_2(X)$ which allows us to focus on  identification only in terms of $1-H_2(X)$. 
Because a claim is filed only if it involves a damage above the deductible, 
we identify the truncated damage distributions  
\begin{eqnarray*}
H^*_c(\cdot|X)\equiv \frac{H(\cdot|X)-H_c(X)}{1-H_c(X))},
\end{eqnarray*}
on $[dd_c(X),\overline{d}(X)]$  from the subpopulation of insurees buying the coverage $(t_c(X),dd_c(X))$ for $c=1,2$.
Differentiating the above equations and taking their ratio show that 
\begin{eqnarray}
\lambda(X)\equiv \frac{h_2^*(D|X)}{h_1^*(D|X)}=\frac{1-H_1(X)}{1-H_2(X)}, 
\end{eqnarray}  
for all  $D\geq dd_1(X)$, where $0< \lambda(X) < 1$.
In particular, the function $\lambda(\cdot)$, which is the ratio of the truncated damage densities, is identified  from the data, while  $H(\cdot|X)$ is identified on $[dd_2(X),\overline{d}(X)]$
up to the knowledge of $H_2(X)$.  

We follow similar steps as in  Case 3 with $\tilde{\theta}\equiv (1-H_2(X))\theta$ replacing $\theta$ while modifying the argument as $J$ is unobserved. 
To identify the marginal density $f_{\tilde{\theta}|X}(\cdot|\cdot)$ of $\tilde{\theta}$ given $X$,  we exploit  the observed number of reported accidents $J^*_c$. Using a similar argument as in (20), the moment generating function of $J^*$ given $(\chi,X)$, where $\chi\in\{1,2\}$ indicates the insuree's contract choice, is 
\begin{eqnarray}
M_{J^*|\chi,X}(t|c,x) &=& {\rm E}[e^{J^*t}|\chi=c,X=x] \nonumber\\
&=& {\rm E}\{{\rm E}[e^{J^*t}|J,\chi,X]|\chi=c,X=x \}  \nonumber\\
&=&  {\rm E} \left\{[H_\chi(X) + (1-H_\chi(X))e^{t}]^J|\chi=c, X=x \right\} \nonumber\\
   &=& {\rm E}\left\{ {\rm E}[e^{J \log [H_\chi(X)+(1-H_\chi(X))e^t]}|\theta,\chi,X] |\chi=c,X=x   \right\} \nonumber\\
  &=&  {\rm E} \left[e^{\theta[H_\chi(X) + (1-H_\chi(X))e^t-1]}|\chi=c,X=x  \right] \nonumber\\
  &=& M_{\theta|\chi,X} [(1-H_\chi(X))(e^t-1)|c,x],
\end{eqnarray}
where the third equality uses the moment generating function of $J^*$ given $(J,\chi,X)$, which is  distributed as a Binomial  ${\cal B}(J,1-H_\chi(X))$ 
using A3-(ii),  and the fifth equality 
follows from A3-(iii) and  the moment generating function of the Poisson distribution. 
Thus, 
\begin{eqnarray*}
M_{\theta|\chi,X}[u|c,x] = M_{J^*|\chi,X} \left[\log \left(1 + \frac{u}{1-H_\chi(X)}   \right)\Big{|}c,x   \right],
\end{eqnarray*}
for $u\in (-1+H_\chi(X),+\infty)$.
In particular, the distribution of risk $\theta$ given $(\chi,X)$ is identified up to the knowledge of $H_\chi(X)$. 

Since $\tilde{\theta}=(1-H_2(X)) \theta$, its moment generating function given $(\chi,X)$ is 
\begin{eqnarray}
M_{\tilde{\theta}|\chi,X}(u|c,x) &=& M_{\theta|\chi,X}(u(1-H_2(x))|c,x) \nonumber \nonumber\\
&=& \left\{ \begin{array}{ll} M_{J^*|\chi,X}\left[\log \left(1 + \frac{u}{\lambda(x)} \right)|1,x \right]  & \ {\rm if} \ c=1, \\
                              M_{J^*|\chi,X}\left[\log \left(1 + u \right)|2,x \right]                   &\ {\rm if} \ c=2,
\end{array} \right. 
\end{eqnarray}
for all $u \in (-\lambda(x),+\infty)$ and $u\in (-1,+\infty)$, respectively.
Thus, the moment generating function of $\tilde{\theta}$ given $X$ is
\begin{eqnarray}
M_{\tilde{\theta}|X}(u|x) &=& {\rm E} \{{\rm E}[e^{u \tilde{\theta}}|\chi,X]|X=x \}\nonumber \\
  &=& M_{J^*|\chi,X} \left[\log \left(1\!+\! \frac{u}{\lambda(x)}\right) |1,x  \right]\! \nu_1(x) \nonumber\\     
    &&   
        + M_{J^*|\chi,X} \left[\log \left(1\!+\! u \right) |2,x  \right]\! \nu_2(x),
\end{eqnarray}
for $u \in (-\lambda(x),+\infty)$, showing that $f_{\tilde{\theta}|X}(\cdot|\cdot)$ is identified as $\lambda(X)$, $\nu_1(X)$ and $\nu_2(X)$ are known from the data. Since $f_{\theta|X}(\theta|x)= (1-H_2(x)) f_{\tilde{\theta}|X}((1-H_2(x))\theta|X)$,
the former density is identified up to $H_2(x)$.
 
In the second step, as in Case 3, we consider  the probability that an insuree with risk $\theta$ and characteristics $X$ chooses the coverage $(t_1(X),dd_1(X))$.
Using (13) and $1-H(D|X)=(1-H_2(X))(1-H^*_2(D|X))$, we remark that the optimal frontier between buying the two coverages in the space $(\tilde{\theta},a)$ is given by
\begin{eqnarray}
\tilde{\theta}(a,X) = \frac{t_2(X)-t_1(X)}{\int_{dd_2(X)}^{dd_1(X)} e^{aD} [1-H^*_2(D|X)] dD},
\end{eqnarray} 
leading to the inverse $a(\tilde{\theta},X)$, which is identified. As before, from Bayes' rule we have 
\begin{eqnarray}
F_{a|\tilde{\theta},X}(a(\tilde{\theta},x)|\tilde{\theta},x) = \frac{f_{\tilde{\theta}|\chi,X}(\tilde{\theta}|1,x) \nu_1(x) }
         {f_{\tilde{\theta}|X}(\tilde{\theta}|x)},
\end{eqnarray}
where $\nu_1(x)$  and  $f_{\tilde{\theta}|X}(\tilde{\theta}|x)$ are identified. 
Moreover, $f_{\tilde{\theta}|\chi,X} (\cdot|1,x)$  is identified because its moment generating function
$M_{\tilde{\theta}|\chi,X}(\cdot|1,x)$ is identified on $(-\lambda(x,),+\infty)$ as shown above.

In the third step, we note that $F_{a|\tilde{\theta},X}(a(\tilde{\theta},x)|\tilde{\theta},x)= F_{a|\theta,X}(a(\theta,x)|\theta,x)$
thereby identifying the latter up to $H_2(x)$ since $\tilde{\theta}= (1-H_2(x))\theta$.
Under A4, the rest of the argument is similar as in Case 3 leading to the  identification of $F_{a|\theta,W}(\cdot|\cdot,\cdot)$ and then of $F(\theta,a|W,Z)$   up to the knowledge of $H_2(X)$.
We have then proved the following result.

\medskip\noindent
{\bf Proposition 4:} {\em   Suppose that two insurance coverages are offered to each insuree and  accidents are observed only when damages are above the deductible. Under A3 and A4, the structure $[F(\cdot,\cdot|X),H(\cdot|X)]$ is identified up to $H_2(X)$.}

\medskip
Up to now, we have not used the optimality of the offered coverages.
Specifically, we have not used the FOC (14)--(18) determining the optimal insurance coverages $(t_1(X),$ $dd_1(X),t_2(X),dd_2(X))$.
One might ask whether the use of these FOC may help in identifying  some features of the structure or even the full structure itself.
For instance, we note that (18) identifies $\underline{a}(X)$ because the latter solves the identifying equation
\begin{eqnarray*}
t_1(X)  = \frac{\underline{\tilde{\theta}}(X)}{\underline{a}(X)} \int_{dd_1(X)}^{\overline{d}(X)} \left(e^{\underline{a}(X) D} - e^{\underline{a}(X) dd_1(X)}\right) h_2^*(D|X)dD, 
\end{eqnarray*}
using $h(D|X)= [1-H_2(X)] h^*_2(D|X)$ and $\underline{\tilde{\theta}}(X)= \underline{\theta}(X)[1-H_2(X)]$. 
A consequence of Proposition 4 is that the structure $[F(\cdot,\cdot|X), H(\cdot|X)]$ is identified if and only if $H_2(X)$ is identified.
The next lemma shows that $H_2(X)$ is not identified even when considering coverage optimality through the 
FOC (14)-(18).

\medskip\noindent
{\bf Lemma 5:} {\em Suppose that two insurance coverages are offered to each insuree and  accidents are observed only when damages are above the deductible. Under A3 and A4, $H_2(X)$ is not identified.}

\medskip
The proof is given in the appendix. It relies on exhibiting  an observationally equivalent structure.
The nonidentification may be surprising but can be explained as follows. It arises from a compensation between the increase (decrease) in the number of accidents and an appropriate decrease (increase) in the probability of damages being greater than the deductible.
From the insuree's perspective, such a compensation maintains the relative ranking between the two contracts. Thus, if a  $(\theta,a)$-insuree 
buys $(t_1(X),dd_1(X))$ then  the $((1-H_2(X))\theta,a)$-insuree  also buys the same coverage if there is an appropriate increase  in the probability of damages being greater than $dd_1(X)$.
From the insurer's perspective,  the decrease  in the average number of accidents is compensated by an appropriate increase in the probability that the damage is above the deductible. Thus the expected payment to the insuree  remains the same under either coverage.

\section{Discussion and Model Restrictions} 

This section discusses identification strategies for the probability $H_2(X)$ and characterizes all the model restrictions on observables associated with the model of Case 4.

\subsection{Identification Strategies for $H_2(X)$}

From Section  4.2, any  assumption that identifies $H_2(X)$  identifies the structure $[F(\cdot,\cdot|X),$ $H(\cdot|X)]$ on its support. 
We discuss some identifying assumptions/conditions for $H_2(X)$ as well as its partial identification.
A first  strategy to identify $H_2(X)$ is to parameterize the damage distribution $H(\cdot|X)$ as $H(\cdot|X;\beta)$ on $[0,\overline{d}(X)]$ with $\beta \in {\cal B} \subset \Real^q$.  Observations on reported damages $D^*$ identify $\beta$ and hence $H(\cdot|X)$ on $[0, \overline{d}(X)]$.
Thus $H_2(X)\equiv H(dd_2(X)|X;\beta)$ is identified.
In particular, we can choose a parametrization to fit the estimated truncated damage distribution $H^*(\cdot|X)$.

A second strategy is to consider additional data sources  on the average of either the number of accidents or the damages.
For instance,   suppose that for every $x \in {\cal S}_X$, we know the average number of  accidents   $\mu(x)\equiv {\rm E}[J|X=x]={\rm E}\{{\rm E}[J|\theta,X=x]|X=x\}={\rm E}[\theta|X=x]$ by A3-(iii). For 
 the average number of reported accidents, we have  $\mu^*_c(x)\equiv {\rm E}[J^*|\chi=c, X=x]={\rm E}\{ {\rm E}[J^*|J,\chi=c,X=x]|\chi=c,X=x \}=
{\rm E}[J (1-H_c(X))|\chi=c,X=x]=[1-H_c(x)]
{\rm E}[\theta|\chi=c,X=x]$ for $c=1,2$ since $J^*$ given $(J,\chi,X)$ is distributed as a Binomial with parameters $(J,1-H_{\chi}(X))$.
Thus
\begin{eqnarray*}
\mu(x)&=&  \nu_1(x) {\rm E}[\theta|\chi=1,X=x] + \nu_2(x) {\rm E}[\theta|\chi=2,X=x]\\ 
        &=&\frac{1}{1-H_2(x)} \left( \nu_1(x) \frac{\mu^*_1(x)}{\lambda(x)} + \nu_2(x) \mu^*_2(x) \right) .
\end{eqnarray*}
This leads to the  identification of $H_2(x)$ given that $\nu_c(x)$, $\mu^*_c(x), c=1,2$ and $\lambda(x)$ are identified from the data
as shown in Section 4.2. 
Alternatively, suppose that we know only ${\rm E}[J|X=x_0]$ for some $x_0$.
Using the same argument establishes the identification of $H_2(x_0)$. This  combined with a support assumption such as
$\overline{\theta}(x)=\overline{\theta}$ for every $x$ identifies $H_2(x)$. Specifically, note that we have
$\overline{\tilde{\theta}}(x) = (1-H_2(x)) \overline{\theta}(x)$, where $\overline{\tilde{\theta}}(x)$ is the upper boundary 
of the support of $f_{\tilde{\theta}|X}(\cdot|X=x)$, which is identified as shown in Section 4.2. Applying this equation at $x_0$ identifies 
$\overline{\theta}$ by $\overline{\tilde{\theta}}(x_0)/(1-H_2(x_0))$. Applying again this equation at different values $x$ identifies $H_2(x)$.
A similar argument applies at the lower bound $\underline{\theta}(x)=\underline{\theta}$.
 
Regarding damages, we note that
\begin{eqnarray*}
{\rm E}(D|X=x) =  H_2(x){\rm E}[D|D \leq dd_2(x),X=x]  
    + (1-H_2(x)) {\rm E}[D|D \geq dd_2(x),X=x] ,
\end{eqnarray*}
where ${\rm E}[D|D \geq dd_2(x),X=x]$ is identified from the data.
Thus, for every $x$ it is straightforward to see that identification of $H_2(x)$ requires to know both ${\rm E}[D|D \leq dd_2(x),X=x]$ and
${\rm E}(D|X=x)$. In particular, the knowledge of the latter is not sufficient, in contrast to the previous case in which  the average number of accidents was sufficient for identification.  As above, if one knows ${\rm E}[D|D \leq dd_2(x_0),X=x_0]$ and
${\rm E}(D|X=x_0)$ for some $x_0$ and if either $\overline{\theta}(x)$ or $\underline{\theta}(x)$ is independent of $x$, then $H_2(x)$ is identified for every $x$.

A third strategy is to derive some bounds on the probability $H_2(X)$.
This approach also known as partial identification was popularized by Manski and Tamer (2002) and Chernozhukov, Hong and Tamer (2007). See also Haile and Tamer (2003) and Kovchegov and Yildiz (2009) for nonparametric bounds. Our bounds are in the spirit of the latter as they are nonparametric.
Let  $[F^0(\cdot,\cdot|X), H^0(\cdot|X)]$ be the true structure. 
Given an arbitrary value $x$, Proposition 4 implies that it is sufficient to determine the identified set for $H_2^0(x)$, i.e., the set of values $H_2(x)$ that are observationally equivalent to $H_2^0(x)$.\footnote{To be precise, this is the set of values $H_2(x)$ corresponding to  structures 
$[F(\cdot,\cdot|X),H(\cdot|X)]$ that are observationally equivalent to $[F^0(\cdot,\cdot|X), H^0(\cdot|X)]$.} 
The proof of Lemma 5 shows that any value $H_2(x) = 1 - (1/\kappa)[1-H^0_2(x)]$ for $\kappa > \sup_{\tilde{x}} [1-H^0_2(\tilde{x})]$ is observationally 
equivalent to $H^0_2(x)$. Thus, the identified set for $H^0_2(x)$ contains the interval 
\begin{eqnarray*}
\left(1- \frac{1-H_2^0(x)}{\sup_{\tilde{x}} [1-H^0_2(\tilde{x})]}, 1\right).
\end{eqnarray*} 
For values $x$ for which $1-H_2^0(x)$ is close to the supremum, the left boundary   approaches zero. Hence, the identified set is close to $(0,1)$, which is not informative.

To tighten these bounds, we may rely on
some empirical evidence in Cohen and Einav (2007).
In particular, their estimated damage density  decreases when the damage approaches the deductible from above  suggesting that the density below the deductible is not greater than its value at the deductible. Thus we can assume that  the damage density satisfies
$h(D|x)\leq h[dd_2(x)|x]$ 
for every $D\leq dd_2(x)$ and $x\in {\cal S}_{X}$.
Integrating both sides from $0$ to $dd_2(x)$ we obtain $0\leq H_2(x) \leq dd_2(x) h(dd_2(x)|x)$. 
Dividing both sides  by $1-H_2(x)$, and using the definition of the truncated density $h_2^*(\cdot|x)$, we obtain
\begin{eqnarray*} 
0\leq \frac{H_2(x)}{1-H_2(x)}\leq dd_2(x) h_2^*(dd_2(x)|x).
\end{eqnarray*}
Solving for $H_2(x)$ gives the bounds
\begin{eqnarray*} 
0\leq H_2(x)\leq \frac{dd_2(x)h_2^*(dd_2(x)|x)}{1+dd_2(x)h_2^*(dd_2(x)|x)}\equiv \overline{B}(x).
\end{eqnarray*}
In particular, the upper bound  for $H_2(x)$ is strictly less than 1.  Moreover,
a useful feature of this upper bound is that it can be estimated as it depends on observables.\footnote{Similarly, exploiting  the relationship  $1-H_2(x)=[1-H_1(x)]/\lambda(x)$ we obtain  
\begin{eqnarray*}
1-\lambda(x)\leq H_1(x)\leq 1-\frac{\lambda(x)}{1+dd_2(x)h_2^*(dd_2(x)|x)}.
\end{eqnarray*}
The lower and upper bounds for $H_1(x)$ are strictly larger than zero and smaller than one, respectively.}

\subsection{Model Restrictions}

This section derives the restrictions imposed by the model on observables under the data scenario of Case 4, i.e., a finite number of contracts and a truncated damage distribution.
We can use these restrictions to test the model and its assumptions.
For every insuree, we observe $[J^*,D^*_1,\ldots,D^*_{J^*},\chi, T,DD,X]$, where $D^*_j$ denotes the damage for the $j$th reported accident
and $(T,DD)$ are the premium and deductible chosen by the insuree. From the model, $T$ and $DD$ are given by $T=t_\chi(X)$ and $DD=dd_\chi(X)$, where
$t_\chi(X)$ and  $dd_\chi(X)$ for $\chi=1,2$ are  functions of $X$ satisfying the first-order conditions (14)-(18). Thus, the vector 
of observables has a joint distribution $\Psi(\cdot,\ldots,\cdot)$ with a  density $\psi(\cdot,\ldots,\cdot)= \psi_{D^*_1,\ldots,D^*_{J^*}|J^*,\chi,X}(\cdot,\ldots,\cdot|\cdot,\cdot,\cdot)
\times \psi_{J^*|\chi,X}(\cdot|\cdot,\cdot) \times \psi_{\chi|X}(\cdot|\cdot) \times \psi_{X}(\cdot)$.
 
The next lemma provides necessary and sufficient conditions on the joint distribution $\Psi(\cdot,\ldots,\cdot)$  to be rationalized 
by a structure $[F(\cdot,\cdot|\cdot),H(\cdot|\cdot)] \in {\cal F}_{X} \times {\cal H}_{X}$.
Let ${\cal H}^*_{c X}$ be defined as  the set  ${\cal H}_{X}$  in Definition 2 with the difference that the support is $[dd_c(X),\overline{d}(X)]$ for $c=1,2$. 
We introduce the remaining notations to write the model restrictions implied by the full support assumption and the first-order conditions (14)--(18). The insurer's expected payment per accident given the coverage $c$ and characteristics $x$ is denoted ${\rm E}[P|c,x]= 
\int_{dd_c(x)}^{\overline{d}(x)} (1-\Psi_{D^*|\chi,X}(D|c,x))dD$ for $c=1,2$.
Let  $\tilde{\theta}(a) \equiv \tilde{\theta}(a,x)$ and $a(\theta) \equiv \tilde{\theta}^{-1}(\tilde{\theta},x)$ as in (25) with $H^*_2(D|X)=\Psi_{D^*|\chi,X}(D|2,X)$.  In particular, $\tilde{\theta}(\cdot)$ and $a(\cdot)$ are known from $\Psi(\cdot,\ldots,\cdot)$. Let 
$f_{\tilde{\theta}|\chi,X}(\cdot|\cdot,\cdot)$ and $f_{\tilde{\theta}|X}(\cdot|\cdot)$ be the densities  given by the moment generating functions  (23) and (24) with 
$\nu_c(x)=\psi_{\chi|X}(c|x)$ for $c=1,2$ and $\lambda(x)= \psi_{D^*|\chi,X}(\cdot|2,x)/\psi_{D^*|\chi,X}(\cdot|1,x)$.  These densities are also known from $\Psi(\cdot,\ldots,\cdot)$.
We denote by $\tilde{\underline{\theta}}\equiv\tilde{\underline{\theta}}(x)$ the lower bound of the support of $f_{\tilde{\theta}|X}(\cdot|\cdot)$. 
Let  $f_{\tilde{\theta},a|X}(\cdot,\cdot|\cdot)=  f_{a|\tilde{\theta},X}(\cdot|\cdot,\cdot)  f_{\tilde{\theta}|X}(\cdot|\cdot)$,
where $f_{a|\tilde{\theta},X}(\cdot|\cdot,\cdot)$ is obtained from (26) using A4. 
Let $[\underline{a},\overline{a}]\equiv [\underline{a}(x),\overline{a}(x)]$ be the support of $f_{a|X}(\cdot|x)$, while 
$a^*\equiv a^*(x)= \min \{\overline{a},a(\tilde{\underline{\theta}},x)\}$. Lastly, 
we define
\begin{eqnarray*}
\rho(x) &=& \psi_{\chi,X}(1,x)+\int_{\underline{a}}^{a^*}\!\left[t_{1}(x)\!-\!\tilde{\theta}(a)  {\rm E}[P|1,x] \right]
f_{\tilde{\theta},a|X}(\tilde{\theta}(a),a|x)\frac{\partial \tilde{\theta}(a)}{\partial t_1} da  \nonumber \\
&& -\int_{a^*}^{\overline{a}}\!\left[t_2(x)\!-\!\tilde{\theta}(a)  {\rm E}[P|2,x]  \right]f_{\tilde{\theta},a|X}(\tilde{\theta}(a),a|x)\frac{\partial \tilde{\theta}(a)}{\partial t_1} d a,
\end{eqnarray*}
which expresses the Lagrange multiplier in terms of observables using (14).

\medskip\noindent
{\bf Lemma 6 (Rationalization Lemma):} {\em 
Let $\Psi(\cdot,\ldots,\cdot)$ be the distribution  of $(J^*,D^*_1,\ldots,$ $D^*_{J^*},$ $\chi,X)$.
Under A3 and A4, $[F(\cdot,\cdot|\cdot),H(\cdot|\cdot)]\in{\cal F}_{X} \times {\cal H}_{X}$ rationalizes $\Psi(\cdot,\ldots,\cdot)$ if and only if the latter satisfies  the following conditions:

\noindent
(i)  $\Psi_{D^*_1, \ldots, D^*_{J^*}|J^*,\chi,X}(\cdot,\ldots,\cdot|\cdot,\cdot,\cdot)=\prod_{j=1}^{J^*} \Psi_{D^*_j|\chi,X}(\cdot|\cdot,\cdot)$, where $\Psi_{D^*_j|\chi,X}(\cdot|\cdot,\cdot)=\Psi_{D^*|\chi,X}(\cdot|\cdot,\cdot)$ $\in {\cal H}^*_{\chi X}$, 

\noindent
(ii) For all $x \in {\cal S}_{X}$, $\psi_{D^*|\chi,X}(\cdot|2,x)$ and $\psi_{D^*|\chi,X}(\cdot|1,x)$ are strictly positive on $[dd_2(x),\overline{d}(x)]$ and $[dd_1(x), \overline{d}(x)]$, respectively. Moreover, their ratio $\lambda(x)$ is independent of
$d\in[dd_1(x),$   $\overline{d}(x)]$ with $0<\lambda(x)<1$,

\noindent
(iii)  For every $(\tilde{\theta},x) \in {\cal S}_{\tilde{\theta}X}$
\begin{eqnarray*} 
\left\{ \frac{f_{\tilde{\theta}|\chi,W,Z}[\tilde{\theta}|1,w,z] \psi_{\chi|W,Z}(1|w,z)}{
f_{\tilde{\theta}|W,Z}(\tilde{\theta}|w,z)}; z \in {\cal S}_{Z|\tilde{\theta}w}  \right\} =[0,1],
\end{eqnarray*}

\noindent
(iv) 
The coverage terms
$t_1(\cdot),t_2(\cdot),$ $ dd_1(\cdot), dd_2(\cdot)$ satisfy $0< t_1(\cdot)<t_2(\cdot),$ $\overline{d}(\cdot) > dd_1(\cdot)$ $> dd_2(\cdot)> 0$, and
\begin{eqnarray}
&&\!\!\!\!\!\!\int_{\underline{a}}^{a^*}\!\left[t_1(x)\!-\!\tilde{\theta}(a)  {\rm E}[P|1,x]   \right]f_{\tilde{\theta},a|X}(\tilde{\theta}(a),a|x)\frac{\partial \tilde{\theta}(a)}{\partial dd_1} da +  
{\rm E}[J^*|1,x]\psi_{\chi,X,Z}(1,x)  \nonumber\\
&&\!\!\!\!\! -\int_{{a}^*}^{\overline{a}}\!\left[t_2(x)\!-\!\tilde{\theta}(a)    {\rm E}[P|2,x] \right] f_{\tilde{\theta},a|X}(\tilde{\theta}(a),a|x)
\frac{\partial \tilde{\theta}(a)}{\partial dd_1} da   -\!\rho(x)\underline{\tilde{\theta}}e^{\underline{a}dd_1(x)}=0\\
&&\!\!\!\!\!\!    \int_{\underline{a}}^{a^*}\!\left[t_1(x)\!-\!\tilde{\theta}(a)  {\rm E}[P|1,x]  \right]f_{\tilde{\theta},a|X}(\tilde{\theta}(a),a|x)
\frac{\partial \tilde{\theta}(a)}{\partial t_2} da  +  \psi_{\chi|X}(2|x) \nonumber \\
&& \!\!\!\!\! -\int_{a^*}^{\overline{a}}\!\left[t_2(x)\!-\!\tilde{\theta}(a)
 {\rm E}[P|2,x]    \right]f_{\tilde{\theta},a|X}(\tilde{\theta}(a),a|x)
\frac{\partial \tilde{\theta}(a)}{\partial t_2}da=0\\
&&\!\!\!\!\!\!\int_{\underline{a}}^{a^*}\!\left[t_1(x)\!-\!\tilde{\theta}(a)  {\rm E}[P|1,x]   \right]f_{\tilde{\theta},a|X}(\tilde{\theta}(a),a|x)
\frac{\partial \tilde{\theta}(a)}{\partial dd_2}da +    {\rm E}(J^*|\chi=2,x)   \psi_{\chi,X}(2|x)  \nonumber\\
&& \!\!\!\!\!-\int_{a^*}^{\overline{a}}\!\left[t_2(x)\!-\!\tilde{\theta}(a)  {\rm E}[P|2,x]   \right]f_{\tilde{\theta},a|X}(\tilde{\theta}(a),a|x)\frac{\partial \tilde{\theta}(a)}{\partial dd_2} da=0\\
&&\!\!\!\!\!\!t_1(x)=\frac{\underline{\tilde{\theta}}}{\underline{a}}\!\left[\int_{dd_1(x)}^{\overline{d}(x)}\left(e^{\underline{a}D}-e^{\underline{a}dd_1(x)}\right)\psi_{D^*|\chi,X}(D|1,x) dD\right].
\end{eqnarray}
}

\medskip
Condition (i) says that reported damages are independent and identically distributed given the coverage choice and individual characteristics.
In addition, reported damages are independent of the reported number of accidents given these variables. This is a  consequence of A3-(i, ii) on damages and number of accidents.
Condition (ii) requires that the densities of reported damages, given coverage choice and individual characteristics, are strictly positive on their supports.
More importantly, the ratio of these densities needs to be independent of the level of reported damage following (21). This property is also a consequence of A3-(i, ii), i.e.,  damages are i.i.d  and independent from the  coverage choice and hence from 
$(\theta,a)$.
Condition (iii) says that the probability   for choosing coverage 1 by a
$(\theta,a)$-insuree   takes all values in $[0,1]$  as the characteristic $Z$ varies. This follows from (26) and the full support condition in A4-(ii).
Condition (iv)  relates  the distribution of observables  to the coverage terms.  In particular, it requires that the optimal premium and deductible for the two coverages must satisfy the FOC (14)-(18). 
There is also a fifth condition  that follows from the 
compact support of the joint distribution of risk and risk aversion and 
its non-vanishing density in Definition 1. This technical condition is given in the Appendix.

The rationalization lemma
is important for several reasons. First, the insurance model with multidimensional private information does impose some 
restrictions on observables. In view of bunching due to multidimensional screening, and a finite number of coverages, 
one could have expected otherwise. For instance, in auction models, a restriction arises from the monotonicity of the equilibrium 
bidding strategy, which is not present here because of the finite number of contracts.
Second,  Lemma 6 characterizes all the restrictions on the distribution of observables that we  can use to test the validity of the model and its assumptions. Violation  of a single restriction by the data would reject the model.
We can then develop some testing procedures for each condition.
For instance, we can test (i) using conditional independence tests. See (say)  Su and White (2008). 
We can test the independence of $\lambda(x)$ from  damage by noting that the ratio of the densities is  equal to 
$\psi_{D^*|\chi,X}(dd_1(x)|2,x)/\psi_{D^*|\chi,X}(dd_1(x)|1,x)$. We can then derive a Cram\'er-von Mises type test
relying on nonparametric estimates of the densities following Brown and Wegkamp (2002).
Condition (iii) implies that the full support assumption in A4 is also testable.

Third, (iv)  provides  restrictions on the coverage terms suggesting that we can test their optimality. This contrasts with the previous structural literature in which one  assumes that the observations are the outcomes of some equilibrium.  For instance, in auctions, identification relies on the optimality of observed bids. This represents a strong assumption that  might be questionable from an empirical point of view.  When the number of contracts is finite,  we do not use optimality of the coverage terms   to identify the model structure. 
Thus, we can use (27)--(30) to test the optimality of the observed coverages $(T_1,DD_1,T_2,DD_2)$ in the case of a monopoly. From an empirical point of view,  the system (27)-(30) gives
the optimal coverages from observables. Hence,  it allows  us to assess the profit  loss  for the insurer from using the actual coverages. 
Fourth,  because restrictions (i)--(iii) do not  require that the insurer is a monopoly, they are also valid to test Assumptions A3 and A4 under alternative forms of competition in the insurance industry.

\section{Conclusion}

Our paper addresses the identification of  insurance models with multidimensional screening, where insurees have private information about both their risk and risk aversion.  Our model also includes a random damage and  the possibility of multiple accidents. Screening of insurees relies on their certainty equivalence. Specifically, we investigate how data availability on the number of offered coverages and reported accidents affects identification of the model primitives through several data scenarios. Overall, the number of accidents plays a crucial role and we  identify the model structure despite  bunching due to  multidimensional screening and/or the finite number of offered coverages. In particular,  our identification results under a finite number of coverages apply to any form of competition. Specifically, they identify the distribution of inusrees' risk and risk aversion for each firm in the industry.
In addition, we provide all the restrictions imposed by the model on observables. An interesting feature  is that  optimality of the offered finite coverages can be tested separately as identification of the model does not rely on this property. 

In terms of future lines of research, first   our results extend to a broad range of insurance data such as in health provided the analyst observes a repeated outcome, e.g. insurees' claims.   In particular, we may want to extend our identification results when damages are no longer mutually independent and  correlated with insuree's private information to allow for moral hazard. Second, in the case of automobile insurance, we could endogenize
the car choice given insuree's risk and risk aversion. This would lead to  a model explaining the car choice, the coverage choice, the number of accidents and the damages.  
Third, our identification results are constructive and thus provide  explicit equations for developing a nonparametric estimation procedure.   Our model restrictions 
can  be used to develop a test of the model validity and of the coverage optimality.  These restrictions are also  the basis for testing adverse selection in insurance within a multidimensional private information setting. 
Several existing data sets  on automobile and/or home insurance used in Israel (2005a,b),  Cohen and Einav (2007), Sydnor (2010) and Barseghyan, Molinari, O'Donoghue and Teitelbaum (2013) can be reanalyzed in view of our results.

\newpage

\begin{center}
{\bf Appendix}
\end{center}

\small

\setcounter{equation}{0}
\renewcommand{\theequation}{A.\arabic{equation}}

\medskip\noindent
{\bf Proof of Lemma 1:}  The derivatives of the certainty equivalences (5) and (6) with respect to $\theta$ give $-(\phi_a-1)/a$ and $-(\phi^*_a-1)/a$, respectively. Since $\phi_a >1$ and $\phi^*_a>1$, we obtain the desired result.
Regarding the derivative of (5) with respect to $a$, we obtain
$$
\frac{\partial  CE(0,0;\theta,a)}{\partial a}= - \theta \left[ \frac{a {\rm E}[D\exp(aD)]- {\rm E}[\exp(aD)]+1}{a^2}\right].
$$
It suffices to show that the numerator in brackets is positive. It is equal to ${\rm E}[aD \exp(aD)- \exp(aD)+1]$. Let $\tilde{X}=aD$, it is easy to show that $\tilde{X} \exp(\tilde{X}) - \exp(\tilde{X}) +1$ is an increasing function equal to 0 at $\tilde{X}=0$. Since $aD\geq 0$,  the numerator is positive and hence the derivative is negative. 
A similar argument applies to $CE(t,dd;\theta,a)$
by letting $\tilde{X}= a \ {\rm min}(dd,D)$. $\Box$

\medskip\noindent
{\bf Derivation of First-Order Conditions (11) and (12):} The Hamiltonian is
\begin{eqnarray*}
H(t(s),dd(s))&=&\left[t(s)-{\rm E}(\theta|s)\int_{dd(s)}^{\overline{d}}(1-H(D))dD\right]k(s) \\   
            &&+v(s)  t'(s)+ y(s)  dd'(s) +r(s)\left[ dd'(s) +\eta(s, a(s),dd(s)) t'(s) \right],
\end{eqnarray*}
where $t(s)$ and $dd(s)$ are the state variables, $t'(s)$ and $dd'(s)$ are the control variables, $v(s)$, $y(s)$ and $r(s)$ are the co-state 
variables.  The first-order conditions are
\begin{eqnarray*}
&&\frac{\partial H}{\partial t'(s)}=  v(s) + r(s) \eta(s,a(s),dd(s))=0 \\
&&\frac{\partial H}{\partial dd'(s)}= y(s) + r(s)=0 \\
&&- \frac{\partial H}{\partial t} = - k(s) = v'(s) \\
&&-\frac{\partial H}{\partial dd}= - \left[{\rm E}[\theta|s] (1-H(dd)) k(s) + r(s)
\frac{\partial \eta(s,a(s),dd(s))}{\partial dd} t'(s)\right]= y'(s)
\end{eqnarray*}
with transversality conditions $y(\underline{s})=0$ and $v(\underline{s})=0$. Integrating the third equation and using the transversality condition $v(\underline{s})$=0 gives $-K(s)=v(s)$. The first two equations give $K(s)-y(s) \eta[s,a(s),dd(s)]=0$.
Using $r(s)=-y(s)$ and  (8) in rewriting the last equation give the desired result. $\Box$

\medskip\noindent
{\bf Proof of Lemma 2:} Let $s'>s$ and $\theta$ be fixed and arbitrary. Following (6), the certainty equivalence when buying insurance can be written as
\begin{eqnarray*}
CE(t(s),dd(s);\theta,a)= w-t(s)- m(dd(s),s),
\end{eqnarray*}
where $m(dd(s),s)=(\theta/a)  [\int_0^{dd(s)} e^{aD} dH(D)+ e^{a dd(s)}(1-H(dd(s)))-1]$ and $(\theta,a)$ is such  that $s(\theta,a)=s$.
The (IC) constraints for $s$ and $s'$ give
\begin{eqnarray*}
&& w-t(s) -m(dd(s),s) \geq w-t(s')-m(dd(s'),s)\\
&& w-t(s')-m(dd(s'),s') \geq w-t(s)-m(dd(s),s').
\end{eqnarray*}
Adding the two inequalities give upon simplification
\begin{eqnarray*}
m(dd(s'),s)-m(dd(s),s) \geq m(dd(s'),s')-m(dd(s),s').
\end{eqnarray*}
Since $m(\cdot,\cdot)$ is differentiable in both arguments, we get
\begin{eqnarray}
&&\int_{dd(s)}^{dd(s')} \frac{\partial m(\xi,s)}{\partial \xi} d\xi \geq 
\int_{dd(s)}^{dd(s')} \frac{\partial m(\xi,s')}{\partial \xi} d\xi \nonumber\\
&& \int_{dd(s)}^{dd(s')} \left[\frac{\partial m(\xi,s)}{\partial \xi}-
\frac{\partial m(\xi,s')}{\partial \xi}   \right] d\xi \geq 0 \nonumber\\
&& \int_{dd(s)}^{dd(s')} \int_{s'}^{s} \frac{\partial^2 m(\xi,y)}{\partial \xi \partial y} dy d\xi \geq 0.
\end{eqnarray}
Differentiating $m(\xi,y)$ with respect to $\xi$ gives
\begin{eqnarray*}
\frac{\partial m(\xi,y)}{\partial \xi}=\theta e^{a\xi} (1-H(\xi)).
\end{eqnarray*}
Because $\theta$ is fixed and $s(\theta,a)=y$, then differentiating with respect with $y$ using $a(y)$ gives
\begin{eqnarray*}
\frac{\partial^2 m(\xi,y)}{\partial \xi \partial y}= \theta a'(y) \xi e^{a(y)\xi}
(1-H(\xi)) \leq 0,
\end{eqnarray*}
since $a(\cdot)$ is decreasing in $s$ by Lemma 1. 
Thus, the inner integration in (A.1) is positive. Hence (A.1) holds if and only if 
$dd(s') \geq dd(s)$.
$\Box$

\medskip\noindent
{\bf Proof of Lemma 5:} In view of Proposition 4, $H_2(X)$ is identified if and only if the structure $[F(\cdot,\cdot|X),H(\cdot|X)]$ is. Thus, it suffices to show that the latter is not identified.  Let  $[F(\cdot,\cdot|X),H(\cdot|X)]$ be a structure  satisfying Definitions 1 and 2 as well as A3 and A4.
We construct  a second structure $[\tilde{F}(\cdot,\cdot|X), \tilde{H}(\cdot|X)]$ as follows. Let $\tilde{\theta}=\kappa \theta$ with $\kappa>
\sup_{x \in {\cal S}_{X}} [1-H_2(x)] \geq 0$, while $\tilde{a}=a$ so that $\tilde{f}(\cdot,\cdot|X)= (1/\kappa) f(\cdot/\kappa,\cdot|X)$. 
Let $\tilde{h}(\cdot|X)$ be a strictly positive conditional density on its support $[0,\overline{d}(X)]$ with 
$\tilde{h}(D|X)= (1/\kappa) h(D|X)$ for $D \geq dd_2(X)$. 
Because $0 < \int_{dd_2(x)}^{\overline{d}(x)} \tilde{h}(D|x) dD <1$, it follows that $\kappa > 1-H_2(x)$ for all $x \in {\cal S}_{X}$
as required above.
The second structure $[\tilde{F}(\cdot,\cdot|X), \tilde{H}(\cdot|X)]$ satisfies 
Definitions 1 and 2 as well as A3 and A4  as $\tilde{\theta}(a,X)=\kappa \theta(a,X)$. 

We now show that these two structures are observationally equivalent, i.e. they lead to the same distribution for the observables 
$( J^*, D^*_1,\ldots,D^*_{J^*}, \chi, t_1, dd_1, t_2, dd_2)$ given $X$, where $J^*$ and $D^*$ refer to the number of reported accidents
and their corresponding damages, respectively, while $\chi$ indicates which coverage is chosen by the insuree.
First, we note that the coverage terms are deterministic functions of $X$ solving the FOC (14)--(18).
Thus, from  (25) the optimal frontier for the second structure must be
\begin{eqnarray*} 
\tilde{\theta}(a,X)&=&  \frac{t_2(X)-t_1(X)}{ \int_{dd_2(X)}^{dd_1(X)} e^{aD} (1-\tilde{H}(D|X)) dD} 
                  = \frac{t_2(X)-t_1(X)}{ \int_{dd_2(X)}^{dd_1(X)} e^{aD} \frac{1}{\kappa}(1- H(D|X)) dD}\\
                  &=& \kappa \theta(a,X),
\end{eqnarray*}
thereby showing that the highest risk aversion in $\tilde{{\cal A}}_1$ is $\tilde{a}^*(X)=a^*(X)$. 

Regarding the distribution  $\tilde{\chi}$ given $X$, we note that $\tilde{\chi}=\chi$. The latter follows from  $\tilde{\chi}=1$ if and only if 
$(\tilde{\theta},a) \in \tilde{{\cal A}}_1(X)$, i.e. $\tilde{\theta} \leq \tilde{\theta}(a,X)$ and $\underline{a}(X) \leq a \leq \tilde{a}^*(X)$.
Since $\tilde{\theta}=\kappa \theta$, $ \tilde{\theta}(a,X)=\kappa \theta(a,X)$ and $\tilde{a}^*(X)=a^*(X)$, we have $\tilde{\chi}=1$ if and only if $\chi=1$. Thus, the distributions of $\tilde{\chi}$ and $\xi$ given $X$ are the same, i.e. 
$\tilde{\nu}_c(X)= \nu_c(X)$ for $c=1,2$.
Regarding the distribution of $\tilde{J}^*$ given $(\tilde{\chi},\!X)\!=\!(\chi,\!X)$, from (22) its moment generating function is
\begin{eqnarray*}
M_{\tilde{\theta}|\chi,X} [(1-\tilde{H}_\chi(X))(e^t-1)|c,x] &=& M_{\theta|\chi,X}[(1-H_\chi(X))(e^t-1)|c,x] \\
                                                                 &=& M_{J^*|\chi,X}[t|c,x]
\end{eqnarray*}   
using $1-\tilde{H}_c(X)= (1-H_c(X))/\kappa$, and $M_{\tilde{\theta}|\chi,X}(u|c,x)= M_{\theta|\chi,X}(\kappa u|c,x)$.
Hence, the distribution of $\tilde{J}^*$ given $(\chi,X)$ is the same as that of $J^*$ given $(\chi,X)$.
Regarding the distribution of reported damage $\tilde{D}^*$ given $(\tilde{J}^*,\chi,X)$ is 
\begin{eqnarray*}
\tilde{H}^*_\chi(\cdot|X) = \frac{\tilde{H}(\cdot|X) - \tilde{H}_\chi(X)}{1-\tilde{H}_\chi(X)} = 
\frac{H(\cdot|X) - H_\chi(X)}{1-  H_\chi(X)} = H^*_\chi(\cdot|X)
\end{eqnarray*} 
using $1-\tilde{H}_\chi(\cdot|X)= (1-H_\chi(\cdot|X))/\kappa$. 

Lastly, it remains to show that $(t_1(X), dd_1(X), t_2(X), dd_2(X))$ satisfies the FOC (14)--(18) associated with the second structure.
Using $\tilde{\theta}(a,X)=\kappa \theta(a,X)$, $\tilde{f}(\tilde{\theta}(a,X),a|X)= f(\tilde{\theta}(a,X)/\kappa,a|$ $X)/\kappa= f(\theta(a,X),a|X)/\kappa$, $1-\tilde{H}(D|X)=(1-H(D|X))/\kappa$,
$\tilde{\nu}_c=\nu_c$ and ${\rm E}[\tilde{\theta}|\tilde{{\cal A}}_c]= \kappa {\rm E}[\theta|{\cal A}_c]$,
it can be easily verified that $(t_1(X), dd_1(X), t_2(X), dd_2$ $(X))$ satisfies (14)--(18) with $\tilde{\rho}=\rho$ as soon as 
(14)--(18) hold for the original structure.
Hence, the two structures lead to the same distributions for the observables as desired.$\Box$

\medskip\noindent
{\bf Additional Condition in Lemma 6:}

\noindent
{\em (v)  For $c=1,2$ and all $x \in {\cal S}_{X}$, $\psi_{J^*|\chi,X}(\cdot|c,x)>0$ on $\Integer$ with  a moment generating function defined on $\Real$ such that the right-hand sides of (23) are the moment generating functions of absolutely continuous distributions with densities bounded away from zero on their supports $[\tilde{\underline{\theta}}(1,x),\tilde{\overline{\theta}}(1,x)]$ and $[\tilde{\underline{\theta}}(2,x),\tilde{\overline{\theta}}(2,x)]$ with union equal to $[\tilde{\underline{\theta}}(1,x),\tilde{\overline{\theta}}(2,x)]$ included in $\Real_{++}$.
Moreover, ${\cal S}_{a|\tilde{\theta} w}\equiv \{a: \exists z\in{\cal S}_{Z|\tilde{\theta} w}, a=\tilde{a}(\tilde{\theta},w,z)\}$ is a compact interval in $\Real_{++}$ independent of $\tilde{\theta}$.}

\medskip
Condition (v) states that the support of the distribution of reported accidents, given coverage choice and individual characteristics, 
is the set of integers. The remaining part of (v) follows from the compact support of  $F(\theta,a|X)$, and its non-vanishing density. 
 The conditions on the moment generating function of $J^*$ given $(\chi,X)$  can be replaced by conditions on its characteristic function $\phi_{J^*|\chi,X}(\cdot|c,x)$. Specifically, $\phi_{J^*|\chi,X}(\cdot|c,x)$ is an entire characteristic function such that the right-hand sides of (23)  are  characteristic functions corresponding to absolutely continuous distributions with densities bounded away from zero on their supports $[\tilde{\underline{\theta}}(1,x),\tilde{\overline{\theta}}(1,x)]$ and $[\tilde{\underline{\theta}}(2,x),\tilde{\overline{\theta}}(2,x)]$ with union equal to $[\tilde{\underline{\theta}}(1,x),\tilde{\overline{\theta}}(2,x)]$ included in $\Real_{++}$.\footnote{Such conditions can be written equivalently in more testable forms.  For instance, a function is a characteristic function if and only if it satisfies Bochner's Theorem 4.2.2, and it is entire if and only if it satisfies Theorem 7.2.1.  A characteristic function corresponds to a distribution with bounded support in $\Real_{++}$ if and only if it satisfies Theorem 7.2.3 with (7.2.3) strictly positive.    These theorems and equations are from Lukacs (1960). A well-known sufficient condition for a distribution to be absolutely continuous is that its characteristic function is absolutely integrable, while a necessary condition is that the characteristic function vanishes in the tails.  See Billingsley (1995, pp.345-347).}

\medskip\noindent
{\bf Proof of Lemma 6:}  We first prove necessity.  Let $[F(\cdot,\cdot|\cdot),H(\cdot|\cdot)]\in{\cal F}_{X}\times{\cal H}_{X}$ be a structure that rationalizes $\Psi(\cdot,\ldots,\cdot)$ under A3 and A4.  To prove (i) we follow Guerre, Perrigne and Vuong (2000) proof of Theorem 4 (Conditions C1-C2).  From A3-(i,ii), we have $(D_1,\ldots,D_J)$ i.i.d as $H(\cdot|X)$ conditional upon $(J,\theta,a,X)$.  Thus, $J^*$ follows a ${\cal B}[J,1-H_{\chi}(X)]$ given 
$(J,\theta,a,X)$ since an accident is reported if and only if the damage is above the deductible. For any $(d_1,\ldots,d_j)\in \Real_{+}^j$,
\begin{eqnarray*}
\lefteqn{ {\rm Pr}[D^*_1\leq d_1, \ldots , D^*_j \leq d_j, J^*=j|J,\theta,a,X] } \\
&=& \!\!\!\!\!\sum_{1\leq r_1 \neq \ldots \neq r_j \leq J} \!\!\!\!\!\!\!\!
{\rm Pr}[dd_{\chi}(X)\!\leq\! D_{r_1}\!\!\leq\! d_1, \ldots , dd_{\chi}(X)\!\leq\! D_{r_j}\!\leq\!\! d_j, D_r\! <\!\! dd_{\chi}(X), r\! \not\in\!\{r_1,\ldots,r_j\} |J,\theta,a,X] \\
&=& \!\!\!\frac{J!}{j!(J-j)!} {\rm Pr}[dd_{\chi}(X)\!\leq\! D_1\!\!\leq\! d_1, \ldots , dd_{\chi}(X)\!\leq\! D_{j}\!\leq\!\! d_j, D_r\! <\!\! dd_{\chi}(X), r\!=\!j+1,\ldots,J |J,\theta,a,X] \\
&=& \!\!\!\frac{J!}{j!(J-j)!} \left(\prod_{r=1}^j \left[H(d_r|X)-H_{\chi}(X) \right]\right) \left[H_{\chi}(X) \right]^{J-j}
\end{eqnarray*}  
because $(D_1,\ldots,D_J)$ are i.i.d. as $H(\cdot|X)$ given $(J,\theta,a,X)$.  Since $J^*$ is ${\cal B}[J,1-H_{\chi}(X)]$ given $(J,\theta,a,X)$ we obtain
\begin{eqnarray*}
{\rm Pr}[D^*_1\leq d_1, \ldots , D^*_j \leq d_j| J^*=j, J,\theta,a,X]= 
\prod_{r=1}^j \frac{H(d_r|X)-H_{\chi}(X)}{1-H_{\chi}(X)}
\end{eqnarray*}
showing that $(D^*_1,\ldots,D^*_j)$ are i.i.d as $H^*_{\chi}(X)\in {\cal H}^*_{\chi X}$ given $(J^*=j, J,\theta,a,X)$, and hence given $(J^*=j,\chi,X)$.  Thus, (i) holds.

To prove (ii), we note that $\Psi_{D^*|\chi,X}(\cdot|\cdot,\cdot)=H^*_{\chi}(\cdot)\in{\cal H}^*_{\chi X}$ thereby establishing the first part of (ii).  Moreover, $\psi_{D^*|\chi,X}(d|2,x)/\psi_{D^*|\chi,X}(d|1,x)=(1-H_1(x))/(1-H_2(x))\equiv\lambda(x)$, which is independent of $d\in[dd_1(x),\overline{d}(x)]$ and in $(0,1)$.
Regarding (iii),  for every $(\theta,a,w) \in {\cal S}_{\theta a W}$, 
\begin{eqnarray*}
F_{a|\theta,W}(a|\theta,w)&=& F_{a|\theta,W,Z}[a(\theta,w,z)|\theta,w,z] 
= \frac{f_{\theta|\chi,W,Z}(\theta|1,w,z) \psi_{\chi|W,Z}(1|w,z)}
{f_{\theta|W,Z}(\theta|w,z) },\\
&=&  \frac{f_{\tilde{\theta}|\chi,W,Z}(\tilde{\theta}|1,w,z) \psi_{\chi|w,z}(1|w,z)} {f_{\tilde{\theta}|W,Z}(\tilde{\theta}|w,z) },
\end{eqnarray*} 
for some $z \in {\cal S}_{Z|\theta  w}$, and 
where the first equality follows from A4, the second equality from Bayes' rule, and the third equality from $\tilde{\theta}=(1-H_2(X))\theta$.
Because $a$ can be chosen arbitrarily, it follows that the right-hand side takes all values in $[0,1]$.
Regarding (iv), let $\tilde{\theta}= (1-H_2(X)) \theta$. The proof then follows   the last paragraph of the proof of Lemma 5 with $\kappa=1-H_2(X)$.

To prove (v), we note that 
\begin{eqnarray*}
{\rm Pr}[J^*=j^*|\theta,a,X] = \sum_{j=j^*}^\infty {\rm Pr}[J^*=j^*|J=j,\theta,a,X]{\rm Pr}[J=j|\theta,a,X].
\end{eqnarray*}
Thus, $J^*$ given $(\theta,a,X)$ is a mixture of a ${\cal B}[J,1-H_{\chi}(X)]$ with a mixing ${\cal P}(\theta)$ distribution by A3-(iii).  That is, $\Psi_{J^*|\theta,a,X}(\cdot|\theta,a,x)$ is a ${\cal P}[(1-H_\chi(x))\theta]$ distribution.  Hence, $\psi_{J^*|\chi,X}(\cdot|c,x)=
\int_{{\cal A}_c} \Psi_{J^*|\theta,a,X}(\cdot|\theta,a,x) dF(\theta,a|x)$ thereby establishing $\psi_{J^*|\chi,X}(\cdot|c,x)>0$ on $\Integer$ as $F(\cdot,\cdot|\cdot)\in{\cal F}_{X}$.  The moment generating function of $J^*$ given $(\chi,X)$ exists on $\Real$ in view of (22) since the distribution of $\theta$ given $(\chi,X)$ has a bounded support.  The right-hand sides of (23) must be the moment generating functions of absolutely continuous distributions, with densities bounded away from zero on 
their supports $[\tilde{\underline{\theta}}(1,x),\tilde{\overline{\theta}}(1,x)]$ and $[\tilde{\underline{\theta}}(2,x),\tilde{\overline{\theta}}(2,x)]$ with union equal to $[\tilde{\underline{\theta}}(1,x),\tilde{\overline{\theta}}(2,x)]$ included in $\Real_{++}$, because they are the moment generating functions of $\tilde{\theta}=(1-H_2(X))\theta$ given $(c,x)$, which have such properties.

We now turn to sufficiency.  Let the distribution $\Psi(\cdot,\dots,\cdot)$ of $(J^*,D^*_1,\dots,D^*_{J^*},\chi,X)$ and the contract terms $[t_1(\cdot),dd_1(\cdot),t_2(\cdot),dd_2(\cdot)]$ satisfy (i)--(v).  We need to exhibit  a structure $[F(\cdot,\cdot|\cdot),H(\cdot|\cdot)]\in {\cal F}_{X}\times{\cal H}_{X}$ satisfying A3 and A4 that rationalizes $\Psi(\cdot,\dots,\cdot)$ of $(J^*,D^*_1,\dots,D^*_{J^*},$ $\chi,X)$ and 
$[t_1(\cdot),dd_1(\cdot),t_2(\cdot),dd_2(\cdot)]$.  


In view of the identification argument of Section 4.2, we define $H(\cdot|\cdot)$ as follows: For a constant $\kappa\in(0,1)$, let  $H(D|X)=\kappa\psi_{D^*|\chi,X}(D|2,X)+ (1-\kappa)$ when $D\geq dd_2(X)$.  Note that $H(\cdot|X)$ has a strictly positive density on $[dd_2(X),\overline{d}(X)]$ because $\Psi_{D^*|\chi,X}(\cdot|2,X)\in{\cal H}^*_{2X}$.  For $D\in[0,dd_2(X)]$, let $H(\cdot|X)$ be arbitrary as long as it has a strictly positive density on $[0,dd_2(X)]$.  Thus, $H(\cdot|\cdot)\in {\cal H}_{X}$. Note that $\kappa=1-H(dd_2(X)|X)\equiv 1- H_2(X)$ so that $H^*_2(\cdot|X)\equiv [H(\cdot|X)-H_2(X)]/[1-H_2(X)]=\Psi_{D^*|\chi,X}(\cdot|2,X)$ after straightforward algebra.  Moreover, $\psi_{D^*|\chi,X}(D|2,X)=\lambda(X)$ $\psi_{D^*|\chi,X}(D|1,X)$ for $D\geq dd_1(X)$ by (ii) implying $\lambda(X)=1-\Psi_{D^*|\chi,X}[dd_1(X)|2,X]$ by integration, and $H^*_1(\cdot|X)\equiv [H(\cdot|X)-H_1(X)]/[1-H_1(X)]=\Psi_{D^*|\chi X}(\cdot|1,X)$ after some algebra. Thus, $\Psi_{D^*_1,\ldots,D^*_{J^*}|J^*,\chi,X}(\cdot,\dots,\cdot|\cdot,\cdot,\cdot)$ is rationalized given A3 as long as $\chi$ is a  function of $(\theta,a,X)$ as implied by the theoretical model.

To construct $F(\cdot,\cdot|\cdot)$ we follow the identification argument.  Let $f(\theta|c,X)=\kappa f_{\tilde{\theta}|\chi,X}(\kappa\theta|c,X)$ and $f(\theta|X)=\kappa f_{\tilde{\theta}|X}(\kappa\theta|X)$, where 
these densities exist by condition (v).  In particular, $f(\theta|X)$ is strictly positive on its support $[\tilde{\underline{\theta}}(1,x)/\kappa,\tilde{\overline{\theta}}(2,x)/\kappa]\subset\Real_{++}$.
Turning to $F_{a|\theta,W,Z}(\cdot|\cdot,\cdot,\cdot)=F_{a|\theta,W}(\cdot|\cdot,\cdot)$ by A4-(i), we follow (26). For every $(\theta,w) \in{\cal S}_{\theta W}$, let $F_{a|\theta,W}(\cdot|\theta,w)$ have a strictly positive density on its support ${\cal S}_{a|\tilde{\theta} w}\equiv \{a: \exists z\in{\cal S}_{Z|\tilde{\theta} w}, a=\tilde{a}(\tilde{\theta},w,z)\}={\cal S}_{a|\theta w}\equiv \{a: \exists z\in{\cal S}_{Z|\theta w}, a=a(\theta,w,z)  \}$ satisfying
\begin{eqnarray}
F_{a|\theta,W}[a(\theta,w,z)|\theta,w]
= \frac{ f_{\tilde{\theta}|\chi,W,Z}(\tilde{\theta}|1,w,z)\psi(1|w,z)}{ f_{\tilde{\theta}|W,Z}(\tilde{\theta}|w,z)}
\end{eqnarray}
for every $(\theta,w,z)\in {\cal S}_{\theta WZ}$, where $\tilde{\theta}=\kappa\theta$ and $a(\theta,w,z)\equiv\tilde{a}(\kappa\theta,w,z)$.  
By (iii) the right-hand side has the range of $[0,1]$ as $z$ varies in ${\cal S}_{Z|\tilde{\theta} w}$ for every given $(\tilde{\theta},w)\in{\cal S}_{\tilde{\theta} W}$, i.e., for every given $(\theta,w)\in{\cal S}_{\theta W}$.  Thus, for every $(\theta,w)\in{\cal S}_{\theta W}$
and every $a\in{\cal S}_{a|\theta w}$, there exists a $z\in{\cal S}_{Z}$ such that $a=a(\theta,w,z)$, i.e., A4-(ii) is satisfied.  We can now extend $F_{a|\theta ,W}(\cdot|\theta,w)$ over ${\cal S}_{a|\theta w}$ by $F_{a|\theta,W}(a|\theta,w)=F_{a|\theta,W}[a(\theta,w,z)|\theta,w]$ using (A.2). Thus, $F(\cdot,\cdot|\cdot)\in {\cal F}_{X}$ as desired. 

The structure $[F(\cdot,\cdot|\cdot),H(\cdot|\cdot)]$ constructed as above rationalizes $\Psi_{J^*|\chi,X}(\cdot|\cdot,\cdot)$
because of (23) and the uniqueness of the corresponding density. This structure also rationalizes $\Psi_{\chi|X}(\cdot|\cdot)$.
Specifically, by definition we have 
\begin{eqnarray*}
F_{a|\theta,W}(a(\theta,w,z)|\theta,w) = \frac{f_{\theta|\chi,W,Z}(\theta|1,w,z) \nu_1(w,z)}{ f_{\theta|W,Z}(\theta|w,z)} = 
\frac{f_{\tilde{\theta}|\chi,W,Z}(\tilde{\theta}|1,w,z) \nu_1(w,z)}{ f_{\tilde{\theta}|W,Z}(\tilde{\theta}|w,z)}.
\end{eqnarray*}
Using (A.2) shows that $\nu_1(w,z) = \psi_{\chi|W,Z}(1|w,z)$ as desired.
The fact that the structure rationalizes $(t_1(\cdot),dd_1(\cdot),t_2(\cdot), dd_2(\cdot))$ follows the argument of the 
last paragraph of the proof of Lemma 5.
$\Box$

\newpage

\normalsize
\begin{center}
{\bf References}
\end{center}

\smallskip\noindent
{\bf Armstrong, M.} (1996): ``Multiproduct Nonlinear Pricing,''  {\em Econometrica}, 64, 51-75.

\smallskip\noindent
{\bf Arrow, K.} (1963): ``Uncertainty and the Welfare Economics of 
Medical Care,'' {\em American Economics Review}, 53, 841-973.

\smallskip\noindent
{\bf Aryal, G.} (2015): ``Identifying Multidimensional Adverse Selection Models,''  mimeo.

\smallskip\noindent
{\bf Aryal, G.}, {\bf I. Perrigne}, and {\bf Q. Vuong} (2009): ``Nonidentification of Insurance Models with Probability of Accidents,''
mimeo.

\smallskip\noindent
{\bf Athey, S.} and {\bf P. Haile} (2007):  ``Nonparametric Approaches to Auctions,'' in J. Heckman and 
E. Leamer, eds., {\em Handbook of Econometrics, Volume VI}, Amsterdam: North Holland.

\smallskip\noindent
{\bf Barseghyan, L.}, {\bf F. Molinari}, {\bf T. O'Donoghue} and {\bf J. Teitelbaum} (2013): ``The Nature of Risk Preferences: Evidence from Insurance Choices,'' {\em American Economic Review}, 103, 2499-2529.

\smallskip\noindent
{\bf Basov, S.} (2001): ``Hamiltonian Approach to Multi-Dimensional
Screening,'' {\em Journal of Mathematical Economics}, 36, 77-94.

\smallskip\noindent
{\bf Berry, S.} and {\bf P. Haile} (2014): ``Identification in Differentiated Product Markets,'' {\em Econometrica},  82, 1749-1797.

\smallskip\noindent
{\bf Billingsley, P.} (1995): {\em Probability and Measure},  New York: Academic Press, 3rd Edition.

\smallskip\noindent
{\bf Brown, D.} and {\bf M. Wegkamp} (2002): ``Weighted Minimum Mean-Square Distance from Independence Estimation,'' {\em Econometrica}, 70, 2035-2051.

\smallskip\noindent
{\bf Campo, S.}, {\bf E. Guerre}, {\bf I. Perrigne} and {\bf Q. Vuong} (2011): ``Semiparametric
Estimation of First-Price Auctions with Risk Averse Bidders,'' {\em Review of Economic Studies}, 78, 112-147.

\smallskip\noindent
{\bf  Carneiro, P.}, {\bf K. Hansen} and  {\bf J. Heckman} (2003): ``Estimating Distributions of Treatment Effects with an Application to the Returns to Schooling and Effects of Uncertainty on College Choice,'' {\em  International Economic Review}, 44, 361-422.

\smallskip\noindent
{\bf Chernozhukov, V.}, {\bf H. Hong} and {\bf E. Tamer} (2007): ``Estimation and Confidence Regions for Parameter 
Sets in Econometric Models,'' {\em Econometrica}, 75, 1243-1284.

\smallskip\noindent
{\bf Chiappori, P.A.} and {\bf B. Salanie} (2000): ``Testing for Asymmetric Information in Insurance Markets,''
{\em Journal of Political Economy}, 108, 56-78.

\smallskip\noindent
{\bf Chiappori, P.A.}, {\bf B. Jullien}, {\bf B. Salanie} and {\bf F. Salanie}
(2006): ``Asymmetric Information in Insurance: General Testable Implications,'' {\em Rand Journal of Economics}, 37, 783-798.

\smallskip\noindent
{\bf Cohen, A.} and {\bf L. Einav} (2007): ``Estimating Risk Preferences from Deductible Choice,'' 
{\em American Economic Review}, 97, 745-788.

\smallskip\noindent
{\bf Cohen, A.} and {\bf P. Siegelman} (2010): ``Testing for Adverse 
Selection in Insurance Markets,'' {\em Journal of Risk and Insurance},  77, 39-84.

\smallskip\noindent
{\bf Crawford, G.}  and  {\bf M. Shum} (2007): ``Monopoly Quality Choice in Cable Television,'' {\it Journal of Law and
Economics}, 50, 181-209.

\smallskip\noindent
{\bf Cutler, D.},  {\bf A. Finkelstein}, {\bf K. McGarry} (2008): ``Preference Heterogeneity and Insurance Markets: Explaining a Puzzle of Insurance,''  {\em American Economic Review, Papers and Proceedings},  98, 157-162.

\smallskip\noindent
{\bf Dafny, L.S.} (2010): ``Are Health Insurance Markets Competitive?,''
{\em American Economic Review}, 100, 1399-1431.

\smallskip\noindent
{\bf Einav, L.} and {\bf A. Finkelstein} (2011): ``Selection in Insurance Markets: Theory and Empirics in Pictures,'' {\em Journal of Economic Perspectives}, 25, 115-138.

\smallskip\noindent
{\bf Einav, L.}, {\bf A. Finkelstein} and {\bf P. Schrimpf} (2010): ``Optimal Mandates and the Welfare Cost  of Asymmetric Information: Evidence from the U.K. Annuity Market,''
{\em Econometrica}, 78, 1031-1092.

\smallskip\noindent
{\bf Fang, H.}, {\bf M. Keane} and {\bf D. Silverman} (2008): ``Sources of Advantageous Selection: Evidence from the Medigap Insurance Market,'' {\em Journal of Political Economy}, 116, 303-350.

\smallskip\noindent
{\bf Finkelstein, A.} and {\bf K. McGarry} (2006):  ``Multiple Dimensions of Private Information: Evidence from the Long-term Care Insurance Market,'' {\em American Economic Review}, 96, 938-958.

\smallskip\noindent
{\bf Gayle, G.L.} and {\bf R. Miller} (2015): ``Identifying and Testing Models of Managerial Compensation,'' {\em Review of Economic Studies}, 82, 1074-1118.

\smallskip\noindent
{\bf Gollier, C.} and {\bf H. Schlesinger} (1996): ``Arrow's Theorem on the Optimality of Deductibles: A Stochastic Dominance Approach,''
{\em Economic Theory}, 7, 359-363.

\smallskip\noindent
{\bf Guerre, E.}, {\bf I. Perrigne} and {\bf Q. Vuong} (2000): ``Optimal Nonparametric Estimation of First-Price 
Auctions,'' {\em Econometrica}, 68, 525-574.

\smallskip\noindent
{\bf Guerre, E.}, {\bf I. Perrigne} and {\bf Q. Vuong} (2009): ``Nonparametric Identification of Risk Aversion in First-Price Auctions Under Exclusion Restrictions,'' 
{\em Econometrica}, 77, 1193-1227.

\smallskip\noindent
{\bf Haile, P.} and {\bf E. Tamer} (2003): ``Inference with an Incomplete Model of English Auctions,''
{\em Journal of Political Economy}, 111, 1-51.

\smallskip\noindent
{\bf Heckman, J.J.} (2001): ``Essays: Econometrics and Empirical Economics,'' {\em Journal of Econometrics}, 100, 3-5.

\medskip\noindent
{\bf Honka, E.} (2014): ``Quantifying Search and Switching Costs in US Auto Insurance Industry,'' {\em Rand Journal of Economics}, 45, 847-884.

\smallskip\noindent
{\bf Hurwicz, L.} (1950): ``Generalization of the Concept of Identification,'' in T.C. Koopmans, ed., {\em Statistical Inference in Dynamic Economic Models, Volume 10}, Cowles Commission Monograph, New York: Wiley.

\smallskip\noindent
{\bf Koopmans, T.C.} (1949): ``Identification Problems in Economic Model Construction,'' {\em Econometrica}, 
17, 125-144.

\smallskip\noindent
{\bf Kovchegov, Y.} and {\bf N. Yildiz} (2009): ``Inference in Partially Identified Nonparametric Instrumental Variables Models,''  mimeo.

\smallskip\noindent
{\bf Imbens, G.} and {\bf W. Newey} (2009): ``Identification and Estimation of Triangular Simultaneous Equations Models Without Additivity,'' {\em Econometrica},  77, 1481-1512.

\smallskip\noindent
{\bf Israel, M.} (2005a): ``Services as Experience Goods: An Empirical 
Examination of Consumer Learning in Automobile Insurance,''
{\em American Economic Review}, 95, 1444-1463.

\smallskip\noindent
{\bf Israel, M.} (2005b): ``Tenure Dependence in Consumer-Firm Relationships: An Empirical Analysis of Consumer Departures from Automobile Insurance Firms,'' {\em Rand Journal of Economics}, 36, 165-192.

\smallskip\noindent
{\bf Ivaldi, M.} and {\bf D. Martimort} (1994): ``Competition Under Nonlinear Pricing,'' {\em Annales d'Economie et de Statistiques}, 34, 71-114.

\smallskip\noindent
{\bf Laffont, J.J.}, {\bf E. Maskin} and {\bf J.C. Rochet} (1987): ``Optimal Nonlinear Pricing with Two Dimensional Characteristics,''
in T. Groves, R. Radner and S. Reiter, eds.,  {\em  Information, Incentives and Economic Mechanisms, Essays in Honor of Leonid Hurwicz}, University of Minnesota Press.

\smallskip\noindent
{\bf Landsberger, M.} and {\bf I. Meilijson} (1999): ``A General Model of Insurance under Adverse Selection,'' {\em Economic Theory}, 14, 331-352.

\smallskip\noindent
{\bf Leslie, P.} (2004): ``Price Discrimination in Broadway
Theatre,'' {\em Rand Journal of Economics}, 35, 520-541.

\smallskip\noindent
{\bf Lewbel, A.} (2000): ``Semiparametric Qualitative Response Model Estimation with Unknown Heteroscedasticity or Instrumental Variables," {\em Journal of Econometrics}, 97, 145-177.

\smallskip\noindent
{\bf Lewis, T.} and {\bf D. Sappington} (1989): ``Countervailing Incentives in Agency Problems,'' {\em Journal of Economic Theory}, 
49, 294-313.

\smallskip\noindent
{\bf Lu, J.} and {\bf I. Perrigne} (2008): ``Estimating Risk Aversion from Ascending and Sealed-Bid Auctions: The Case of Timber Auction Data,''  
{\em Journal of Applied Econometrics}, 23, 871-896.

\smallskip\noindent
{\bf Lukacs, E.} (1960): {\em Characteristic Functions}, London: Charles Griffin and Company Limited.

\smallskip\noindent
{\bf Luo, Y.}, {\bf I. Perrigne} and {\bf Q. Vuong} (2012): ``Multiproduct Nonlinear Pricing: Mobile Phone Service and SMS,'' mimeo.

\smallskip\noindent
{\bf Luo, Y.}, {\bf I. Perrigne} and {\bf Q. Vuong} (2013): ``A General Framework for Nonlinear Pricing Data,''  mimeo.

\smallskip\noindent
{\bf Luo, Y.}, {\bf I. Perrigne} and {\bf Q. Vuong} (2015): ``Structural Analysis of Nonlinear Pricing,''  Working Paper, CRATE New York University.

\smallskip\noindent
{\bf Manski, C.} and {\bf E. Tamer} (2002): ``Inference on Regressions with Interval Data on a Regressor or 
Outcome,'' {\em Econometrica}, 70, 519-546.

\smallskip\noindent 
{\bf Matzkin, R. L.} (1992): ``Nonparametric and Distribution-Free Estimation of The Binary Threshold Crossing and The Binary Choice Models ,'' {\em Econometrica}, 60, 239-270.

\smallskip\noindent
{\bf Matzkin, R. L.} (1993): ``Nonparametric Identification and Estimation of Polychotomous Choice Models,'' {\em Journal of Econometrics}, 58, 137-168.

\smallskip\noindent
{\bf Matzkin, R. L.} (1994): ``Restrictions of Economic Theory in Nonparametric Methods,''
in R. Engle and D. McFadden, eds., {\em Handbook of Econometrics, Volume IV}, North Holland.

\smallskip\noindent
{\bf Matzkin, R. L.} (2007): ``Nonparametric Identification,'' in J. Heckman and E. Leamer, eds., {\em Handbook of Econometrics, Volume VI}, North Holland.

\smallskip\noindent
{\bf  Perrigne, I.} and {\bf Q. Vuong} (2011): ``Nonparametric Identification of a  Contract Model with Adverse Selection and Moral Hazard,''  {\it Econometrica},  79, 1499-1539.

\smallskip\noindent
{\bf Pioner, H.}  (2007): ``Semiparametric Identification of Multidimensional Screening Models,''  mimeo. 

\smallskip\noindent
{\bf Rao, B.L.S.P.} (1992): {\em Identifiability in Stochastic Models: Characterization of Probability Distributions}, Academic Press: New York.

\smallskip\noindent
{\bf Rochet, J.C.} and {\bf P. Chone} (1998): ``Ironing, Sweeping and Multidimensional Screening,'' {\em Econometrica}, 66, 783-826.

\smallskip\noindent
{\bf  Rochet, J.C.} and {\bf L. Stole} (2003): ``The Economics of Multidimensional Screening," in M. Dewatripont, L.P.  and S.J.Turnovsky, eds., {\em Advances in Economic Theory: Eighth World Congress},  Cambridge University Press. 

\smallskip\noindent
{\bf Rothschild, M.} and {\bf J. Stiglitz} (1976): ``Equilibrium in Competitive Insurance Markets: 
An Essay on the Economics of Imperfect Information,'' {\em Quarterly Journal of Economics},
90, 236-257.

\smallskip\noindent
{\bf Starc, A.} (2014): ``Insurer Pricing and Consumer Welfare: Evidence from Medigap,'' {\em Rand Journal of Economics}, 45, 198-220.

\smallskip\noindent
{\bf Stiglitz, J.} (1977): ``Monopoly, Nonlinear Pricing and Incomplete Information: The Insurance Market,'' {\em Review of Economic Studies}, 
44, 407-430.

\smallskip\noindent
{\bf Su, L.} and {\bf H. White} (2008): ``A Nonparametric Hellinger Metric Test for Conditional Independence,'' {\em Econometric Theory}, 24, 829-864. 

\smallskip\noindent
{\bf Sydnor, J.} (2010): ``(Over)insuring Modest Risks,'' {\em American Economic Journal: Applied Economics}, 2, 177-199.

\smallskip\noindent
{\bf Tirole, J.} (1988): {\it The Theory of Industrial Organization}, MIT Press.

\smallskip\noindent
{\bf Wilson, R.} (1993): {\em Nonlinear Pricing}, Oxford University Press.

\end{document}